\RequirePackage{fix-cm}
\documentclass[smallextended]{svjour3}       
\smartqed  
\usepackage{graphicx}
\usepackage{amsmath,amsfonts,amssymb}
\usepackage{hyperref}
\usepackage{color}
\usepackage{comment}
\usepackage[normalem]{ulem}
\newcommand{\be}{\begin{equation}}
\newcommand{\ee}{\end{equation}}
\definecolor{cyan}{rgb}{0,0.9,0.9}
\definecolor{orange}{rgb}{0.9,0.5,0}
\definecolor{magenta}{rgb}{1,0,1}
\definecolor{purple}{rgb}{0.8,0.4,0.8}
\definecolor{gray}{rgb}{0.8242,0.8242,0.8242}

\def\l{\ell}

\def\p{\partial}

\def\gccm{{\rm g\,cm^{-3}}}
\def\Msun{{\rm M_{\odot}}}
\def\Mo{\Msun}
\def\GMc2{{\rm G M_{\odot} c^{-2}}}

\def\E{\mathcal{E}}
\def\J{\mathcal{J}}

\def\kt2{\kappa^\text{T}_2}
\def\Mmax{M_\text{max}^\text{TOV}}

\def\Mthr{M_\text{pc}}
\def\kthr{k_\text{pc}}

\begin{document}

\title{Neutron Star Merger Remnants
}
\titlerunning{BNS Merger Remnants} 
\author{Sebastiano Bernuzzi}
\authorrunning{S.~Bernuzzi} 
\institute{S. Bernuzzi \at
Theoretisch-Physikalisches Institut\\
Friedrich-Schiller-Universitaet Jena\\
Max Wien Platz 1\\
D-07743 Jena, Germany\\
\email{sebastiano.bernuzzi@uni-jena.de}              
}

\date{Received: date / Accepted: date}

\maketitle

\begin{abstract} 
  Binary neutron star mergers observations are a unique way to 
  constrain fundamental physics and astrophysics at the extreme. 
  The interpretation of gravitational-wave events and their
  electromagnetic counterparts crucially relies on general-relativistic  
  models of the merger remnants.
  Quantitative models can be obtained only by means of 
  numerical relativity simulations in 3+1 dimensions including 
  detailed input physics for the nuclear matter, electromagnetic and weak interactions. 
  This review summarizes the current understanding of merger
  remnants focusing on some of the aspects that are relevant for 
  multimessenger observations.  
\keywords{Binary neutron star \and mergers \and remnants \and gravitational-waves \and numerical relativity}
\end{abstract}

\section{Introduction}
\label{intro}

The gravitational wave GW170817 is compatible with a
binary neutron star (BNS) inspiral of chirp mass $1.186(1)\Msun$ 
\cite{TheLIGOScientific:2017qsa,Abbott:2018wiz,LIGOScientific:2018mvr}.
Significant signal-to-noise ratio (SNR) is found in the frequency range 30-600 Hz, roughly corresponding to the last thousand orbits to merger for an
equal-mass binary with canonical mass $M\sim 2.8\,\Mo$.
The matched-filtering analysis of GW170817 with tidal waveform templates 
provides us with an estimate of the 
reduced tidal parameter that is distributed around
$\tilde{\Lambda}\sim300$ and smaller than ${\sim}800$
\cite{Damour:2012yf,Favata:2013rwa,DelPozzo:2013ala,Abbott:2018exr}. 
LIGO-Virgo searches for short (${\lesssim}1\,$s), intermediate
(${\lesssim}500\,$s) and long (days) postmerger transients from a
neutron star (NS) remnant 
resulted in upper limits of more than one order of magnitude larger than
those predicted by basic models of quasi-periodic sources~\cite{Lai:1994ke,Cutler:2002nw,Corsi:2009jt,DallOsso:2014hpa,Lasky:2015olc}.
Hence, the LIGO-Virgo detectors' sensitivity was not sufficient to detect a
signal from the merger phase and the remnant, which lie in the kiloHertz 
range~\cite{Abbott:2018hgk}. 
A similar conclusion holds for the second BNS event, GW190425, that
was detected at lower SNR than GW170817\cite{Abbott:2020uma}. 
GW190425 has a chirp mass of $1.44(2)\,\Msun$ and it is associated to
the heaviest BNS source known to date~\footnote{A BH-NS source for GW
  has not been firmly excluded, e.g.~\cite{Kyutoku:2020xka}}.

In absence of a GW detection, the merger remnant can be inferred from
the binary properties and from
the interpretation of the electromagnetic counterparts based on the
theoretical predictions given by numerical relativity (NR)
simulations. The latter are the only method available to determine the
merger outcome and to compute the GW signals from the remnants. 
This review summarizes the current understanding of merger remnants as
determined by NR simulations during the last 20 years~\footnote{This
  review reflects the views of the author who aimed at a brief 
  but updated overview avoiding a more complete
  historical perspective.}.
The presented results are key for the interpretation of future observations
of multimessenger signals from BNS mergers.

Current numerical relativity methods applied to quasicircular mergers
allow us to simulate tens of orbits before merger and the early postmerger 
phase for a timescale of several dynamical periods.
Inspiralling NSs are well described by zero-temperature matter in beta-equilibrium with
maximum density about twice the nuclear saturation density
$\rho_{\rm NS}\sim2-3\rho_0$
($\rho_0\simeq2.3\times10^{14}$~$\gccm$). Electromagnetic fields are not
expected to significantly affect the mass dynamics \cite{Liebling:2010bn,Giacomazzo:2010bx}. Thus, general relativistic
simulations with perfect fluid matter are believed to faithfully
model the orbital phase. The inspiral dynamics can be characterized 
in terms of the binary masses (and spins), and the tidal polarizability
parameters, as described in Sec.~\ref{sec:mrg}.
At the end of the inspiral, about 3-4\% of the initial
gravitational mass is radiated in GWs and the binary merges at typical
GW frequencies ${\sim}1-2$~kHz. 

Binary NS mergers result in the formation of a compact central object,
either a NS or a black hole (BH), eventually surrounded by an
accretion disc~\cite{Shibata:1999wm,Shibata:2002jb,Anderson:2007kz,Baiotti:2008ra}.  
The remnant can be characterized in
first approximation by the NS masses and by the softness of the
(unknown) zero-temperature equation of state (EOS), in particular by the maximum
sustainable mass, $\Mmax$ \cite{Hotokezaka:2011dh,Bauswein:2013jpa}.
Binary remnants with total mass significantly larger than $\Mmax$ 
cannot be sustained by the EOS pressure and by the centrifugal support
of their rotations. Thus, the remnant promptly collapses to a BH during its
formation. A precise definition of prompt BH collapse and the
phenomenology inferred from the simulations are discussed in Sec.~\ref{sec:pc}.

If the remnant does not promptly collapse, its early evolution is driven by 
GW emission and characterized by a luminous GW 
transient emitted at frequencies 
${\sim}2-4\,$kHz \cite{Shibata:2002jb,Stergioulas:2011gd,Bauswein:2011tp,Hotokezaka:2013iia,Takami:2014zpa,Bernuzzi:2015rla}. 
Matter in NS remnants is compressed and heated up to extreme densities
and temperatures, and the baryon mass density can reach $\rho_{\rm
  rem}\sim1.5-2\rho_{\rm NS}\sim3-6\rho_0$ and temperatures
$\gtrsim50\,$MeV \cite{Sekiguchi:2011zd,Perego:2019adq}. 
The NS remnant can either collapse to BH after a ``short life'' on
the dynamical timescale determined by its rotational period, or settle to an
axisymmetric equilibrium configuration on longer timescales.
The black holes that can be produced in BNS mergers are discussed in
Sec.~\ref{sec:bh}. 

After the GW-driven, dynamical phase, the angular momentum of the NS remnant at formation
is well above the Keplerian (mass-shedding) limit of an equilibrium
zero-temperature beta-equilibrated rigidly-rotating configuration with the same baryon mass
\cite{Radice:2018xqa}. Also, the remnant has gravitational mass in
excess of those equilibrium configurations.
Thus, it is far from equilibrium and its 
long-term evolution is determined by the energy and
angular momentum evolution due to magnetohydrodynamics and weak
interactions in the fluid, as well as GW 
emission \cite{Fujibayashi:2017xsz,Radice:2018pdn,Kiuchi:2019lls,Ciolfi:2019fie}.
Neutron-star remnants and their evolutionary phases are discussed in 
Sec.~\ref{sec:ns}.

A key dynamical feature for GW counterparts is the formation of
remnant
discs~\cite{Shibata:2003ga,Shibata:2006nm,Kiuchi:2009jt,Rezzolla:2010fd,Radice:2017lry}. Remnant Discs
of masses ${\sim}0.1\,\Msun$ can form if the matter acquires sufficient rotational 
support during merger. The initial composition and
extension of a remnant disc is 
dependent on whether the central object is a NS or a BH. The disc
evolution starts with a  phase of rapid accretion, but is afterwards
determined by a combination of the 
gravitational pull, the neutrino cooling and the expansion due to viscous
processes and magnetic field stresses \cite{Fernandez:2013tya,Metzger:2014ila,Just:2014fka,Siegel:2014ita,Fujibayashi:2017puw,Fujibayashi:2017xsz}.
The properties of remnant discs are discussed in
Sec.~\ref{sec:disc}. 

During merger a mass ${\sim}{10^{-4}-{10^{-2}}}\,\Msun$ of neutron rich material is expelled
on dynamical timescales \cite{Rosswog:1998hy,Hotokezaka:2012ze,Bauswein:2013yna,Wanajo:2014wha,Radice:2018pdn}. The remnant 
can unbind an even larger amount of material by winds powered
by different mechanisms \cite{Lee:2009uc,Fernandez:2015use,Siegel:2017nub,Fujibayashi:2017puw,Fernandez:2018kax}. These ejection mechanisms and NR-based estimates
of ejecta masses and composition are reviewed in Sec.~\ref{sec:ejecta}.
Mass ejecta from mergers are a key astrophysical site for
heavy-element production via the r-process
\cite{Lattimer:1974slx,Eichler:1989ve,Wanajo:2014wha,Rosswog:2015nja,Kasen:2017sxr,Cowan:2019pkx}. 
The observational imprint of r-process element production is the kilonova electromagnetic transient, that was observed for the first time as the counterpart of GW170817. 
Because of their quasi-isotropic character, kilonovae are considered to be the most
promising EM counterpart for future GW events
\cite{Li:1998bw,Metzger:2010sy,Kasen:2013xka,Tanaka:2013ana,Metzger:2019zeh}.  

Appendix~\ref{sec:NRmethods} summarizes the main input physics and
numerical techniques employed for the NR simulations .

Geometric units $G=c=1$ are used if units are not explicitely indicated.

\section{Merger dynamics}
\label{sec:mrg}

\begin{figure}
  \includegraphics[width=\textwidth]{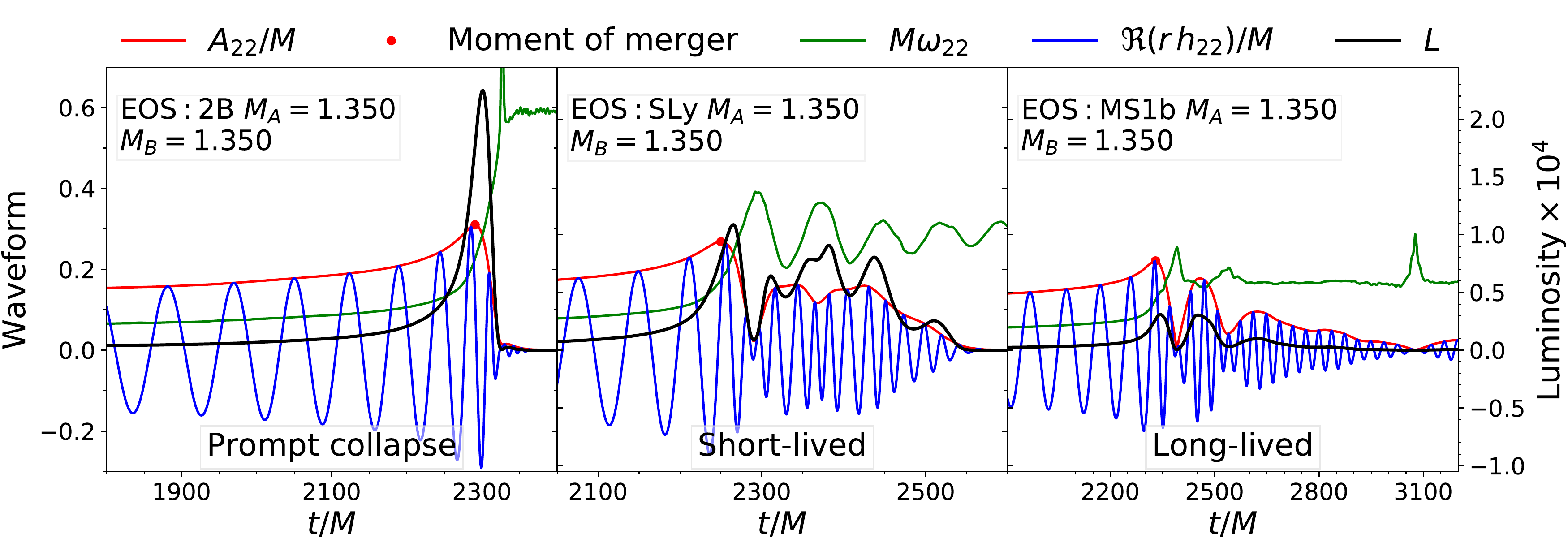}
  \caption{Gravitational-waveforms for representative
    equal-mass irrotational mergers.
    The figure shows the evolution of amplitude, frequency and real
    part of the $(2,2)$ multipole of the GW strain and luminosity. 
    From left to right: prompt collapse, short-lived, and long-lived stable remnant.
    Figure adapted from~\cite{Zappa:2017xba}.}
\label{fig:mrgwf}
\end{figure}

The inspiral BNS dynamics differ from those of the binary black
hole because of the tidal interactions between the NSs. 
Tidal interactions in the post-Newtonian formalism for
self-gravitating and deformable bodies are calculated 
using a multi-chart approach, in which the tidal response of a NS
due to the external gravitational field of the companion (the inner
problem) is matched to an outer problem for the description of the
orbital dynamics and radiation
\cite{Damour:1983a,Damour:1990pi,Damour:1993zn,Racine:2004hs,Damour:2009wj}.
In the local frame of body $A$, the mass 
multipole moments $M^A_L$~\footnote{There exist also spin
  multipole moments and gravitomagnetic tidal moments. The discussion
  here is simplified.} 
are related to the external
gravitoelectric moments $G^A_L$ by 
$M^A_L = \mu^A_\l G_L^A$ 
where $L=i_1...i_\l$ is a multi-index. Analogously to the electric
polarizability of a charge distribution, the coefficients $\mu_\l$
quantify the distortion of the mass distribution due to the external
field. They are often substituted by the dimensionless relativistic Love
numbers obtained by normalizing with the appropriate power of the NS radius, 
\be \label{eq:Lovenum}
k^A_\l := \frac{(2\l-1)!!}{2}\frac{G\mu^A_\l}{R^{2\l+1}_A} \ .
\ee
The practical calculation of the Love numbers reduces to the solution
of stationary perturbations of spherical relativistic stars
\cite{Damour:1983a,Hinderer:2007mb,Damour:2009vw,Binnington:2009bb}.
The Love numbers are thus dependent on
the EOS employed for constructing the equilibrium NS and on the NS compactness,
$C_\text{A}=GM_\text{A}/(c^2R_\text{A})$. 
In the following we will make use exclusively of the quadrupolar
\emph{tidal polarizability
  parameters} defined as~\cite{Flanagan:2007ix,Damour:2009wj} 
\be
\Lambda_A := \frac{2}{3} k^A_2 \left(\frac{GM_A}{R_Ac^2}\right)^{5} \ .
\ee
The two-body relative dynamics in the weak field
regime is described by the Newtonian Hamiltonian with a tidal
term in the potential \cite{Damour:2009wj,Damour:2012yf}
\be
H \simeq 
 \frac{\mu}{2} p^2 + \frac{\mu}{2} \left( -\frac{2GM}{c^2 r} + ... 
- \frac{\kappa_2^T}{r^6} \right) \ ,
\ee
where $\mu$ is the reduced mass of the binary. 
The tidal coupling constant $\kappa_2^T$ is defined as 
\be
 \kappa_2^\text{T} := 
 \dfrac{3}{2} \left[ \Lambda_2^\text{A}
   \left(\frac{M_\text{A}}{M}\right)^4 \frac{M_{B}}{M} +
     (A\leftrightarrow B) \right] \ ,
\ee
and parametrizes at leading
(Newtonian) order the tidal interactions in the binary.
The formula above indicates that tidal interactions are 
attractive and short-range~\footnote{The dependency on the orbital 
  separation follows immediately from the Lagrangian at leading order
  $L\approx +\mu_2 G_{ab}G^{ab}$ and the
  general property $G_L\propto\p_L u \propto r^{\l+1}$.}. 
The effect of tidal interactions in the inspiral is illustrated by 
the (modified) Kepler law \cite{Damour:2009wj},
\be
\Omega^2 r^3 = GM\left[ 1+ 12\frac{M_A}{M_B}\frac{R_A^5}{r^5}k^A_2 +
  (A\leftrightarrow B)\right] \ .
\ee 
At a given radius the frequency is higher if the tidal interactions
are present. Thus, the motion is accelerated by tidal effects and the
system merges earlier and at a lower frequency.  
We shall see that, while the details of tidal interactions during merger can be 
quantified only by general relativistic hydrodynamical simulations,
these basic results are key to characterize the merger data from the
simulations. Note that the reduced tidal parameter \cite{Favata:2013rwa}
\be
\label{eq:LambdaT}
\tilde\Lambda := \frac{16}{13}
\frac{(M_\text{A} + 12 M_\text{B}) M_\text{A}^4}{M^5}\Lambda_\text{A}
+ (A\leftrightarrow B)\ ,
\ee
is also used to parametrize tides at leading order in place of $\kt2$. Although not
the same quantity, $\tilde\Lambda$ and $\kt2$ will be used here for
the same purposes. 
The ranges for BNSs are $\kt2\approx(20,500)$ and
$\tilde\Lambda\approx(50,2000)$.
Softer EOS, larger masses and higher mass-ratios result in smaller
values of $\kt2$ (or $\tilde\Lambda$). 
In what follows we discuss an effective characterization of
the merger properties relevant for the later discussion on the merger
remnant. 

\begin{figure}
  \includegraphics[width=.9\textwidth]{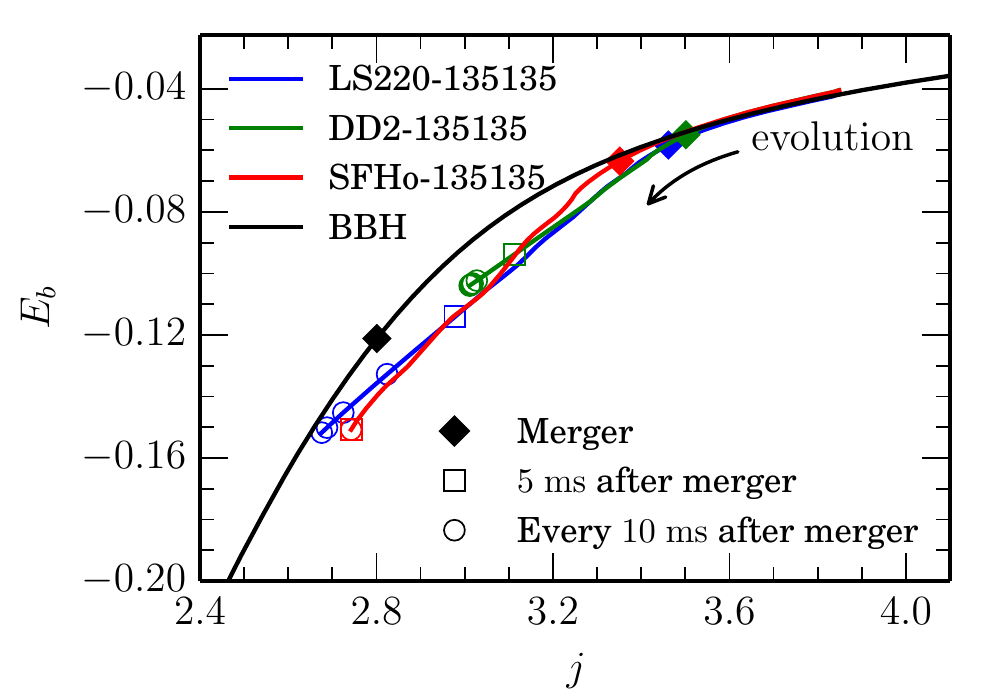}
  \includegraphics[width=.9\textwidth]{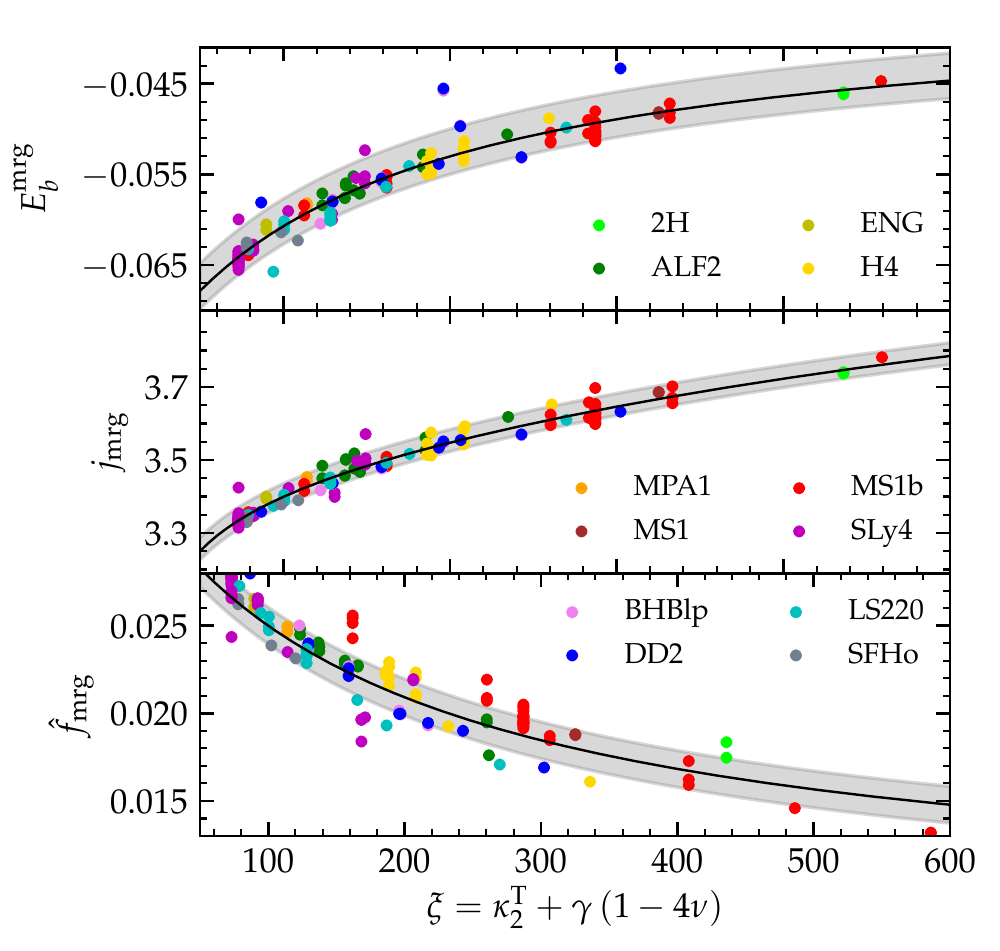}
  \caption{Energetics of BNS mergers and quasiuniversal
    (EOS-independent) relations at
    the moment of merger. 
    Top: Evolution of the reduced binding
    energy and angular momentum (see Eq.~\eqref{eq:Eb}-\ref{eq:j}) for
    representative models and comparison to the binary black hole
    case. The moment of merger is marked by a squared black bullet,
    subsequent times are marked with empty bullets.
    Figure from~\cite{Bernuzzi:2015opx}.
    Bottom: Reduced binding energy, and angular momentum and GW
    frequency at the moment of merger as a function of the
    $\xi(\kt2,\nu)$ parameter (Eq.~\eqref{eq:xi}) from the {\tt CoRe} database of
    simulations. Note that the frequency is rescaled by the mass to give a dimensionless quantity, $\hat{f}_{\rm mrg}=Mf_{\rm mrg}$.
    Gray bands represent the fit 90\% confidence region. 
    Figure adapted from~\cite{Breschi:2019srl}.}
  \label{fig:quasiu}
\end{figure}

The BNS dynamics in numerical relativity are usually 
studied by considering the gravitational radiation computed during the
simulations. The latter is extracted from coordinate spheres at finite
radii $R$ and extrapolated to null-infinity. Simulations
resolve the first modes of the multipolar decomposition, 
\begin{equation}
  R\left(h_+ - i h_\times\right) = \sum_{\ell=2}^{\infty}\sum_{m=-\l}^\l h_{\ell m}(t)\; {}^{-2}Y_{\ell m}(\theta,\varphi)
  \approx h_{22}(t)\; {}^{-2}Y_{22}(\theta,\varphi) + c.c. \ , 
\end{equation}
where ${}^{-2}Y_{\ell m}$ are the spin-weighted $s=-2$ spherical harmonics.
Examples of circular merger gravitational-waves are shown in
Fig.~\ref{fig:mrgwf} together with the instantaneous GW frequency and
luminosity. All the waveforms show the chirp behaviour, 
predicted by the post-Newtonian formalism, that terminates at a characteristic amplitude peak,
the time of which is sometimes referred to as the \emph{moment of
merger} (and distinguished from the merger process). The moment of
merger marks the end of the chirp signal.
Note that the luminosity peak is delayed with respect to the amplitude peak.

A gauge invariant way to characterize the BNS dynamics using
simulation data is to consider the reduced binding energy and angular
momentum of the binary, computed as \cite{Damour:2011fu,Bernuzzi:2012ci} 
\begin{align}
  \label{eq:Eb}
E_b &  =  - \dfrac{M - \Delta E_\text{GW}}{\mu} = 
  \dfrac{\left(M_\text{ADM} - \Delta \E_\text{GW}\right) - M}{\mu} \\
  \label{eq:j}
  j & = \dfrac{J_\text{ADM}-\Delta \J_\text{GW}}{M\mu} \ .
\end{align}

Above, $M$ is the binary mass and 
$\Delta \E_\text{GW}$ and $\Delta \J_\text{GW}$ are the radiated
energy and angular momentum computed from the multipoles
$h_{\l m}$ during a simulation.
The total binding energy $\Delta E_\text{GW}$ and the binary's angular
momentum are computed from $\Delta \E_\text{GW}$ and $\Delta \J_\text{GW}$ by subtracting the contribution of the
Arnowitt-Deser-Misner (ADM) energy and angular momentum of the initial
data. During the evolution, a binary emits energy and angular momentum and both $j$ and $E_b$
decrease from their initial values, as shown in the top panel of
Fig.~\ref{fig:quasiu}. A dynamical frequency can be
defined as
\be\label{eq:Omg}
M\Omega = \frac{\p E_b}{\p j}\,
\ee
using the Hamilton-Jacobi equations. This frequency corresponds to
half the GW frequency $\omega_{22}=-\Im{(\dot{h}_{22}/h_{22})}$
of the dominant $(2,2)$ mode, and thus can be identified with the binary's orbital frequency during the inspiral \cite{Damour:2011fu}.
However, the validity of Eq.~\eqref{eq:Omg} is not restricted to the
inspiral, and $\Omega$ can be used to characterize also the postmerger
evolution. Simulations have shown that the instantaneous frequency of the postmerger waveform is also $\omega_{22}\approx 2\Omega$ \cite{Bernuzzi:2015rla}.  
  
As suggested by Fig.~\ref{fig:mrgwf}, the GW quantities (frequency,
peak amplitude and 
luminosity) and thus also the energetics are very dependent on the binary
mass and mass ratio as well as on the NSs' EOS and spins.
However, using the analytical estimates presented above it is possible
to describe all the numerical data in simple terms.
At sufficiently high frequencies the short range 
tides significantly contribute to the binary interaction
energy~\footnote{%
  A more precise and formal argument is discussed in
  \cite{Bernuzzi:2014kca,Breschi:2019srl}.} and the key dynamical quantities 
and the GW are functions of the tidal parameter \cite{Bernuzzi:2014kca}. 
For example, the properties of every simulated equal-mass binary at the moment of
merger are very well captured by $\kt2$ solely.
The fact that the latter parameter encodes to a very good accuracy the EOS is sometimes referred to as {\it quasiuniversality}; 
relations like $f(\kt2)$ are called EOS-independent or EOS-insensitive
relations.
Mass-ratio effects up to $q=M_A/M_B\sim2$ can be described by further considering the parametrization 
\be\label{eq:xi}
\xi = \kappa_2^T + \gamma (1-4 \nu) \ ,
\ee
where $\nu=\mu M\in[0,1/4]$ and $\gamma$ is a fitting parameter \cite{Breschi:2019srl}.

Figure~\ref{fig:quasiu} shows the robusteness of this description for
a large number of irrotational BNS simulations. 
More compact (small $\kt2$) and more massive binaries emit more
energy, as expected. 
A fiducial equal-mass merger emits about 3-4\% of
the mass in GWs by the end of the chirp phase (for irrotational binaries). 
The angular momentum of the system at merger is
larger the less compact the binary is and the larger (smaller) the mass
ratio $q$ ($\nu$) is. In other terms, binaries with
NSs with large radii merge at larger separations. The GW merger
frequency can be fit to a simple function of $\xi$ 
\be
\label{eq:fmrg.nr}
f^\text{mrg}_\text{GW} \simeq 2.405 \left(\frac{1 + 1.307\, \cdot 10^{-3} \xi}{1 + 5.001\,\cdot 10^{-3} \xi }\right) \left(\frac{M}{2.8 \Msun}\right)\ {\rm kHz} \ ,
\ee
with $\gamma\simeq3200$. Similar relations exists for all the relevant
dynamical quantities, such as the binding energy, the angular
momentum, or the GW luminosity at merger
\cite{Bernuzzi:2014owa,Zappa:2017xba,Breschi:2019srl}. 
The effects due to the NS
rotation can also be included in this picture. The largest spin effect is given by 
spin-orbit interactions that depend, to first approximation, on the magnitude
and sign of the projection of the spin along the orbital angular momentum,
$S_z$. For small spins the effect is linear in spin and, for example, 
the angular momentum at merger is $j_S \approx j_0 \pm S_z/M\mu$
\cite{Bernuzzi:2013rza,Bernuzzi:2014owa}.  

We shall see in the following that $\kt2$ (or $\tilde\Lambda$) is a useful ``order
parameter'' also for some properties of the remnant. 
While there are no binary dynamics in this case, the
remnant quantities at early times are largely determined by the
conditions at merger.

\section{Prompt collapse}
\label{sec:pc}

\begin{figure}
  \includegraphics[width=\textwidth]{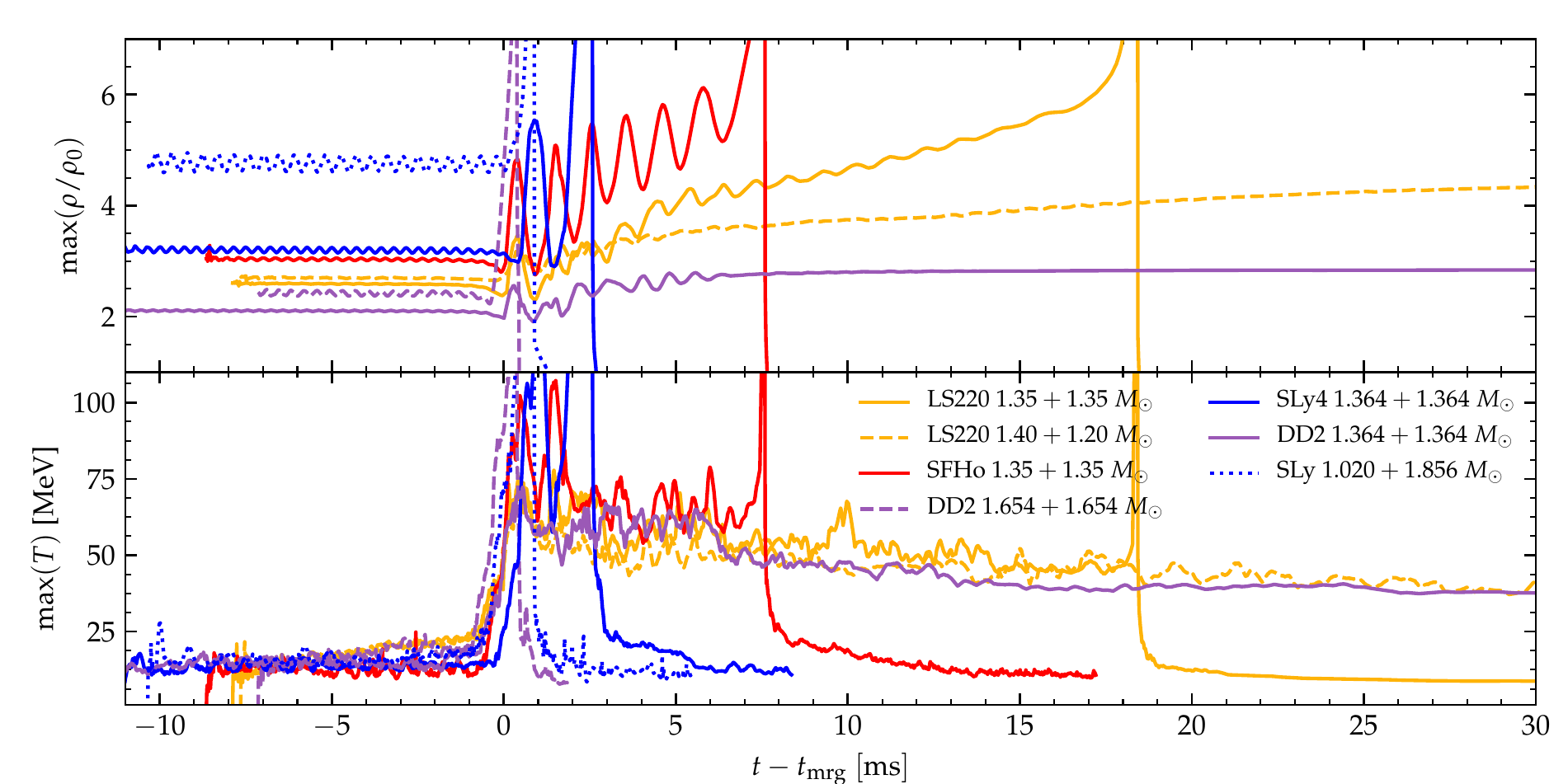}
  \caption{Evolution of the maximum density and temperature during
    mergers of representative binaries.
    The legend indicates the total mass.
    The DD2 $1.654+1.654\ \Msun$ model is an example of prompt collapse;
    SLy $1.020+1.856\ \Msun$ is an example of accretion induced prompt
    collapse. In each case the maximum density monotonically
    increases to the collapse. SLy $1.364+1.364\ \Msun$, SFHo and
    LS220 $1.35+1.35\ \Msun$ are short-lived remnants collapsing to a BH
    at different times. Note that the oscillations in the maximum density  are
    related to the NS cores bouncing and are correlated to the temperature peaks.
    LS220 $1.20+1.40\ \Msun$ shows that moderate mass ratios can 
    increase the lifetime of the remnant (compare to the LS220 equal
    mass case).
    DD2 $1.364+1.364\ \Msun$ is an example of long-lived remnant
    surviving for more than $100$~ms.
    Figure adapted from~\cite{Radice:2018pdn,Perego:2019adq,Bernuzzi:2020txg}.}
\label{fig:rho_t}
\end{figure}

Prompt BH collapse mergers can be defined as those mergers in which the
NS cores collision has no bounce, but instead the remnant immediately collapses 
at formation (See also the discussion in Sec.~\ref{sec:ns}).
Prompt collapse happens within ${\sim}1-2\,$ms from the moment of
merger and can be identified by the maximum density monotonically
increasing to the collapse. Two examples of prompt collapse mergers are 
shown in Fig.~\ref{fig:rho_t}, for which the maximum density increases
beyond $6\rho_0$ at $t\approx t_{\rm mrg}$. This definition of prompt
collapse implies negligible shocked dynamical ejecta because the bulk
of this mass ejection comes precisely from the (first) core bounce
\cite{Radice:2018pdn}; we shall expand on this point in Sec.~\ref{sec:ejecta}.
An example of a merger waveform for a prompt collapse is shown in the
first panel of Fig.~\ref{fig:mrgwf}. The signal's amplitude goes
rapidly to zero while the GW frequency increases to the BH
quasi-normal-mode frequencies. At this point in the simulation, an
apparent horizon has formed and it has reached an approximately
stationary state. 

Numerical-relativity simulations predict that circular and
equal-mass BNS mergers will be followed by a prompt collapse to a BH,
if the total gravitational mass $M$ of the 
binary exceeds a threshold mass, given by \cite{Hotokezaka:2011dh,Bauswein:2013jpa} 
\be\label{eq:mpc}
\Mthr = \kthr \Mmax \ , 
\ee
where $\kthr$ depends on the unknown EOS. 
Current simulations of irrotational, $q\sim1$ BNSs spanning a sample of 18 hadronic
EOS and comparable masses find that $1.3 \lesssim \kthr\lesssim 1.7$ 
\cite{Hotokezaka:2011dh,Bauswein:2013jpa,Zappa:2017xba,Dietrich:2018phi,Koppel:2019pys}.
For these data $\kthr$ shows an approximately
EOS-insensitive linear behaviour in the compactness $C$ (or in the
radius) of a reference nonrotating NS at equilibrium \cite{Bauswein:2013jpa}.
For example, using the extended set of data from
\cite{Hotokezaka:2011dh,Bauswein:2013jpa,Koppel:2019pys}, and choosing
the maximum NS compactness as a reference one finds the best fit
\cite{Agathos:2019sah} 
\begin{equation} \label{eq:kpc}
\kthr (C_\text{max}) = -(3.29\pm0.23) \, C_\text{max} + (2.392\pm0.064)  \ .
\end{equation}
Note that under the hypothesis that the merger did not promptly form a
BH, the inversion of Eq.~\eqref{eq:mpc} leads to a bound on
the maximum NS mass. This argument can be 
used to estimate the maximum NS mass after GW170817, by interpreting
the GW counterpart as evidence for a NS
remnant~\cite{Margalit:2017dij,Shibata:2017xdx,Rezzolla:2017aly,Ruiz:2017due}.

An alternative model for the prompt collapse threshold based on NR
data is based on the tidal parameter $\kt2$ \cite{Zappa:2017xba}. An 
analysis of comparable-mass data of the {\tt CoRe} collaboration
finds that all the reported prompt collapse mergers are captured by the
condition~\footnote{%
  The corresponding value in $\tilde\Lambda$ is
  \be\label{eq:Lam_thr}
  338 \lesssim \tilde\Lambda_\text{pc} \lesssim 386 \ .
  \ee
} 
\be\label{eq:kappapc}
\kt2 < \kappa^{\rm T}_\text{pc} \approx 80\pm40 \ ,
\ee
which is a quasiuniversal relation. For
equal-mass BNSs, $\kt2$ can be interpreted as a measure of the
binary compactness with more compact binaries leading to earlier BH
formation. Note that Eq.~\eqref{eq:kappapc}, differently from
Eq.~\eqref{eq:mpc}, contains a dependence on the mass ratio.
Improved phenomenological descriptions of the collapse threshold can 
be obtained by parametrizing the threshold using both the maximum mass
and the tidal parameter $\kt2$ \cite{Bauswein:2020aag}. 

The above prompt collapse models are valid for comparable masses and
irrotational (no NS spin) mergers. 
For a given total mass, moderate mass ratios can extend
the remnant lifetime with respect to an equal mass BNS because of the
less violent fusion of the NS cores and a partial tidal disruption
that distributes angular momentum at larger radii in the
remnant \cite{Bauswein:2013jpa}.
If the total mass is sufficiently large, the primary NS can be close to
$\Mmax$ and the material accreting from the (partial) tidal disruption of the 
companion can favour a prompt collapse.
Moreover, spin-orbit interactions have repulsive or attractive
character depending on the sign of the spin projection along the
orbital angular momentum (spins aligned or 
antialigned). Hence, they can either increase (or decrease) the
angular momentum support of the remnant and delay (anticipate) BH
collapse \cite{Kastaun:2013mv,Bernuzzi:2013rza,Dietrich:2016lyp}. 

The prompt collapse models above indeed fail for large mass ratios
$q\sim1.5-2$ \cite{Bernuzzi:2020txg}. 
In BNSs with increasing mass ratios and fixed chirp masses, the companion
NS undergoes a progressively more significant tidal disruption.
For a sufficiently soft EOS, the collapse in these mergers is triggered by
the accretion of the companion onto the massive primary star. 
This ``accretion-induced prompt collapse'' scenario should be always
present after a critical mass ratio in connection to the maximum NS
mass. A rough estimate of the threshold is given by modiyfing
Eq.~\eqref{eq:mpc} as $\Mthr(\nu)\sim\Mthr\cdot(4\nu)^{3/5}$, and it
indicates that extreme mass ratios favour prompt collapse. 

A systematic numerical-relativity investigation of the prompt collapse
threshold varying the input EOS models 
(for example also considering hyperons and phase
transitions \cite{Sekiguchi:2011mc,Radice:2016rys,Bauswein:2018bma,Most:2018eaw,Bauswein:2020aag}), 
masses, mass ratio and spin is
presently missing but rather urgent for a quantitative understanding
of the merger dynamics. 
Related to this, it remains challenging to construct an EOS-insensitive 
(universal) relation for robustly determining the prompt collapse from
binary properties.

\section{Remnant black holes}
\label{sec:bh}

Black holes produced by the collapse of irrotational binary merger remnants
(either prompt collapse or short-lived) are found with dimensionless
spin \cite{Kiuchi:2010ze,Kastaun:2013mv,Bernuzzi:2013rza,Bernuzzi:2015opx,Dietrich:2016hky,Bernuzzi:2020txg}
\be
0.6 \lesssim a_\text{BH} \lesssim 0.875 \ .
\ee
This interval can be expected from the merger quasiuniversal relations
presented in Sec.~\ref{sec:mrg}. The relations for $E_b^\text{mrg}(\kt2)$
and $j_\text{mrg}(\kt2)$ at the moment of merger give upper limits for
the BH mass and spin 
\be
M_{\rm BH} < E_b^\text{mrg}\nu M \ \ \mbox{ and } \ \ a_{\rm BH} < j_\text{mrg}\nu \ , \ \
\ee
assuming the remnant would instantaneously collapse to a BH without GW
emission nor remnant disc/ejecta. The reduced angular momentum range
in Fig.~\ref{fig:quasiu} is $3.2\lesssim j_\text{mrg}\lesssim3.8$,
with smaller values obtained for smaller $\kt2$. 
Binaries with $\kt2\gtrsim250$ have stiff EOS and the remnants are
either short or long-lived. Such remnants emit in GWs at least the same amount
of binding energy that they posses at merger (Fig.~\ref{fig:quasiu}, top 
panel), hence one can focus on binaries that collapse promptly with $\kt2<120$
(Eq.~\eqref{eq:kappapc}) and obtain
$a_{\rm BH}<0.875$ for equal-mass BNS ($\nu=1/4$).

The largest BH spins, $a_{\rm BH}\sim0.8$, are obtained for 
equal-mass prompt collapse mergers. Note that, in this case, the postmerger GW luminosity is comparable to that of the
moment of merger and that very light discs are formed.
For large mass ratios the angular momentum at merger is distributed in a
massive accretion disc and the BH spin is below the upper limit.
Black holes formed by the collapse of short-lived NS remnants have
typically smaller spins than those produced in 
prompt collapses (for a given mass), because their postmerger GW emission 
is significant (as will be discussed in Sec.~\ref{sec:ns}) and
they are surrounded by massive accretion discs.

Remnant BHs can spin up due to the disc accretion and, in principle,
can reach almost maximal spins \cite{Bardeen:1970zz,Thorne:1974ve}. 
In practice however, Keplerian discs in merger remnants are too light to
significantly spin-up the BH.  
Moreover, ordered poloidal magnetic fields between the disc and the horizon can
transport angular momentum outward into the bulk of the disc and even
arrest the
accretion~\cite{Gammie:2003qi,Narayan:2003by,Tchekhovskoy:2011a}. The
disc accretion can be further modified by the angular momentum losses
due to winds on the same timescales
\cite{Lee:2005se,Christie:2019lim,Siegel:2017jug}, and the launch of a jet might also spindown the BH \cite{Benson:2009}. 
The evolution of the remnant BH on timescales of seconds is an open
question related to the accretion disc dynamics, that will be further 
discussed in Sec.\ref{sec:disc}. 

The upper limits on the BH rotation inferred from NR simulations should
be considered in models of electromagnetic counterparts. For example,
in short-gamma-ray burst models (SGRBs) the energy deposition by neutrino 
pair-annihilation depends strongly on the BH spin \cite{Zalamea:2010ax}.
For fixed accretion rate, the energy deposition by neutrinos from a
disk accreting onto a BH with $a_{\rm BH}=0.7$ can be up to a factor 100 times
smaller than for a disk feeding a maximally spinning BH. On the other
hand, $a_\text{BH}$ does not significantly constrain SGRB models
invoking magnetic mechanisms, which can easily account for the required
energies even in the absence of a highly spinning BH~\cite{Nakar:2007yr}.
Note that in the Penrose/Blandford-Znajek mechanism the BH rotational energy is
extracted at a rate proportional to $a_\text{BH}^2$ at leading order
in spin~\cite{Blandford:1977ds,Tchekhovskoy:2011a}. 
We refer the reader to recent reviews for a complete discussion on the
accretion flow onto BHs and its connection to SGRBs \cite{Beloborodov:2008nx,Berger:2013jza,Liu:2017kga}.

\section{Remnant neutron stars}
\label{sec:ns}

The observations of pulsars with masses ${\sim}2\Msun$ 
\cite{Antoniadis:2013pzd,Cromartie:2019kug} constrain EOS
models to support maximum masses larger than ${\sim}2\Msun$. In this
scenario, a likely outcome of a fiducial $M\sim2.8\Msun$ merger is a NS 
remnant, e.g. \cite{Hotokezaka:2011dh}. The properties and evolution
of these NS remnants discussed here below are subject of intense research
and closely linked to observations of kiloHertz GW and mergers' counterparts. 

It is customary to define short-lived NS remnants as those 
collapsing on the timescale of their rotational periods (tens
of milliseconds), and long-lived  
remnants those collapsing on significantly longer timescales. Often,
short-lived remnants are referred to
as hypermassive NSs (HMNS), while long-lived remnants are referred to as
either supramassive NSs (SMNS) or massive NSs (MNS).
Throughout this work we do not use the names HMNS and SMNS for merger
remnants~\footnote{Note, however, that the
  nomenclature is retained in some of
  the presented figures.}
 since these names refer to
general-relativistic zero-temperature axisymmetric equilibrium
configurations, but merger remnants are not cold equilibria.
In particular, a HMNS is defined as a
differentially rotating NS at equilbrium with mass 
above the rigidly rotating limit \cite{Baumgarte:1999cq}, while a 
SMNS (MNS) is a rigidly rotating NS at equilibrium with rest mass
larger (smaller) than the nonrotating equilibrium limit $\Mmax$ \cite{Cook:1992,Cook:1993qr}. 

The evolution of the remnant can be approximately separated into an
early (dynamical) GW-driven phase and a secular phase that is
(initially) driven by viscous magnetohydrodynamics processes and neutrino cooling.
The fate of the remnant is determined by a complex interplay of
gravitational, nuclear, weak and electromagnetic interations that often
act on comparable timescales.

\paragraph{Dynamical (GW-driven) phase.}

At formation, NS remnants are very dynamical. The maximum density
and temperature increase immediately after merger as a consequence of
matter compression and the NS cores bounce several times,
e.g. \cite{Perego:2019adq}.  
The more massive and compact the binary, the
faster and the more violent the dynamics are. 
Despite the large relative collision speed, the speed of sound at
densities $\rho \gtrsim \rho_0$ is $c_s \gtrsim 0.2$c and prevents
the formation of hydrodynamical shocks inside the cores. Only at the
surface of the NSs pressure waves can steepen into shock waves which
accelerate matter at the edge of the remnant up to mildly-relativistic
speeds. Thus, matter inside the cores remains
cold ($T \lesssim 10\,{\rm MeV}$) and, while the densest regions of the
cores rotate and fuse, the compressed matter at the contact interface is
pushed outwards. Matter moving outwards reaches temperatures up
to $T\sim 70{-}110~{\rm MeV}$ and forms a pair of co-rotating hot spots
displaced by an angle of ${\sim} \pi/2$ with respect to the densest
cores, e.g. \cite{Kastaun:2016yaf}. The bound matter expelled from the
center forms a disc which is fed by the central remnant with hot and outgoing
density spiral waves streaming from the central region (see also the
discussion in Sec.~\ref{sec:disc}.)

The high temperatures in the remnant determine high neutrino
production and an early burst in neutrino luminosity reaching
${\sim}10^{52-53}$~erg/s, e.g. \cite{Ruffert:1996by,Sekiguchi:2011zd,Palenzuela:2015dqa,Foucart:2015gaa}. 
Simulations including neutrino transport predict the mean neutrino
energies at infinity $E_{\nu_e} (\sim 10~{\rm MeV}) \lesssim
E_{\bar{\nu}_e} (\sim 15~{\rm MeV}) \lesssim  E_{\nu_{\mu,\tau}}
(\sim 20~{\rm MeV})$, with more massive binaries and softer
EOS resulting in higher mean energies
\cite{Sekiguchi:2011zd,Foucart:2016rxm,Endrizzi:2020lwl}. 
Due to the strong dependence of the cross-sections on the incoming
neutrino energy, neutrinos with different energies decouple from
matter in very different regions. 
At the average energies, $\nu_e$ and $\bar{\nu}_e$ decouple at densities between a few
and several times $10^{11}{\rm g~cm^{-3}}$, respectively. Low 
energy neutrinos decouple at around $10^{13} {\rm g~cm^{-3}}$ along
spheroidal neutrino decoupling surfaces \cite{Perego:2014fma,Endrizzi:2020lwl}. 
Because free neutrons are abundant, the
absorption opacities for $\nu_e$ are larger than those for
$\bar{\nu}_e$, while pair processes, responsible for keeping
$\nu_{\mu,\tau}$ and their antiparticles in equilibrium, decouple at
much larger densities and temperatures inside the remnant. 
Electron neutrino and positron
absorption on neutrons increases substantially the electron fraction
in the material, with a larger effect in hotter remnants and along the
polar regions, where neutrino fluxes are more intense due to the lower
optical depths \cite{Wanajo:2014wha,Goriely:2015fqa,Radice:2016dwd,Sekiguchi:2016bjd,Martin:2017dhc,Radice:2018ozg}. 

New degrees of freedom or new matter phases in the EOS at extreme
densities ${\sim}3-5\rho_0$ can also impact the remnant dynamics and leave
detectable imprints on the GW. Examples are matter models
including hyperon production \cite{Sekiguchi:2011mc,Radice:2016rys}
or zero-temperature models of phase transitions to quark-deconfined matter
\cite{Bauswein:2018bma,Most:2018eaw}. In both cases, the EOS models soften
at extreme densities thus favouring more compact remnants and their 
gravitational collapse. The impact of these processes on the dynamics
depends on the densities at which the EOS softens (or stiffens).
Postmerger GWs at kiloHertz frequencies carry, in principle, 
signatures of a rapid EOS softening (or stiffening) at postmerger densities.
However, the unambiguous extraction of information from these
detections will crucially depend on the (unknown) physics details and on the
availability of theoretical models.
For example, if the new matter phases impact
the EOS weakly 
and/or at large densities $\rho\gtrsim5\rho_0$ that are reached
only during the remnant's gravitational collapse, then no significant
imprint in the GW is expected.
In addition, the extraction of information on the EOS or NS properties 
from the kiloHertz spectrum requires the assumption of particular waveform
models that depend on the EOS used in the simulations \cite{Breschi:2019srl}. 
Examples of such models are those connecting the GW spectrum
frequencies to the binary properties, and they are discussed next.

The dynamical phase described above lasts for about
${\sim}10{-}20$~milliseconds until 
the cores have completed their fusion or collapsed to a BH.
During the core fusion, the remnant is a strongly deformed object
with a  pronounced bar-like deformation that powers a significant
emission of GWs. 
The main postmerger GW signature is a short transient with a
spectrum peaking at a few characteristic
frequencies~\cite{Shibata:2002jb,Stergioulas:2011gd,Bauswein:2011tp,Bauswein:2012ya,Hotokezaka:2013iia,Takami:2014zpa,Bernuzzi:2015rla,Radice:2016gym,Lehner:2016lxy,Dietrich:2016hky,Dietrich:2016lyp,Vretinaris:2019spn}. The
main peak frequency is associated with
twice the dynamical frequency of the remnant NS 
at early postmerger times $f_2\sim\Omega/\pi$. It is important to note that $\Omega$ evolves in time and that the GW spectrum is not discrete. The peaks are instead a consequence of the efficiency of the emission process: since the emission is very fast at early times, the spectrum is dominated by the broad peaks at the (approximately constant) frequencies right after merger \cite{Bernuzzi:2015rla,Bernuzzi:2015opx}.
The GW postmerger spectrum can be robustly computed from short and
nonexpensive simulations, thus has been studied in great
detail. The characteristic peaks in the spectrum are often associated
to hydrodynamical modes in the remnant,
e.g.~\cite{Shibata:1999wm,Stergioulas:2011gd,Bernuzzi:2013rza}, and are thus often interpreted in
analogy to linear perturbations of equilibrium NSs 
\cite{Dimmelmeier:2005zk,Passamonti:2007tm,Baiotti:2008nf}.  
We refer to the literature above for detailed analysis of the
characteristic postmerger frequencies and their association to the
hydrodynamical modes in the remnant. 

\begin{figure}
  \centering
  \includegraphics[width=.49\textwidth]{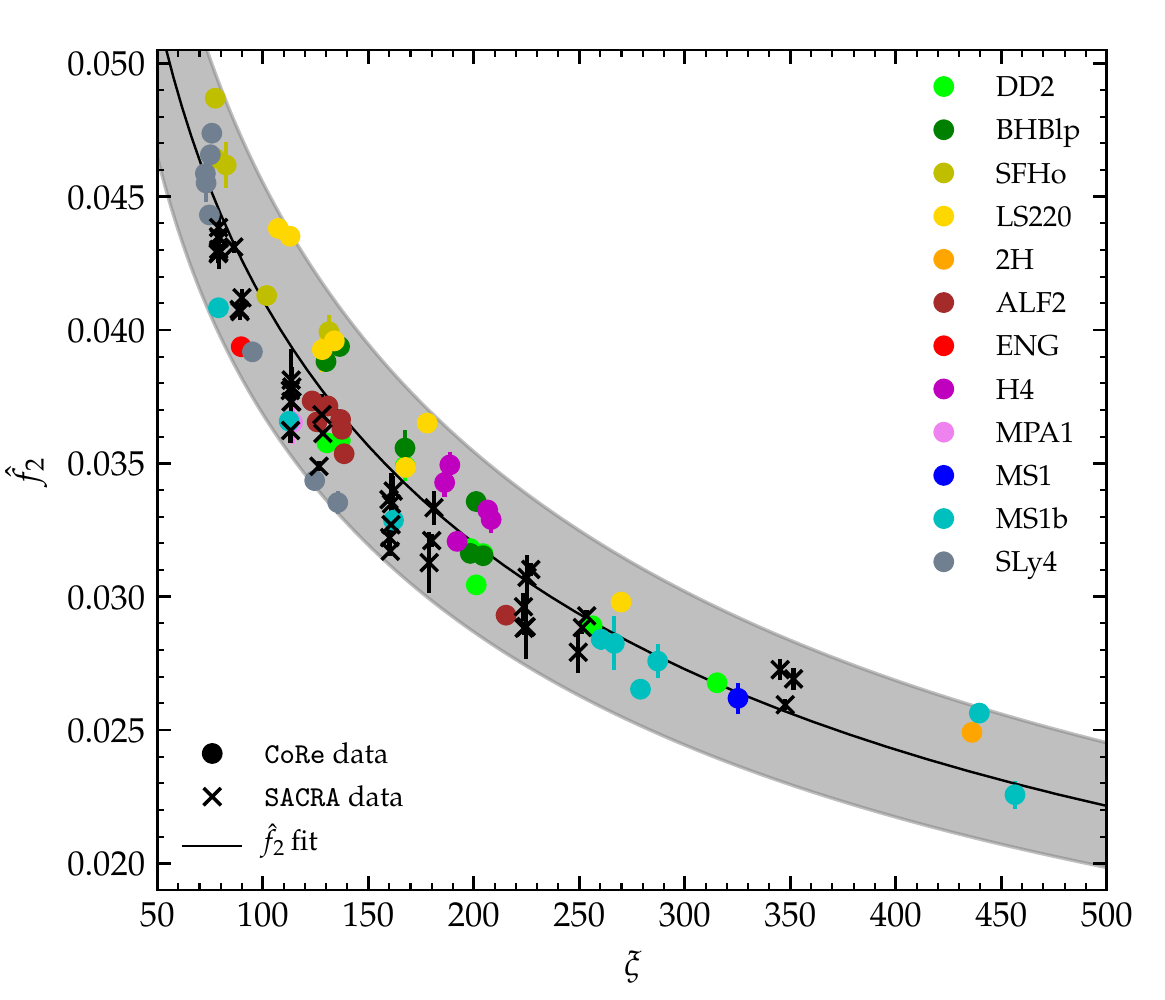}
  \includegraphics[width=.49\textwidth]{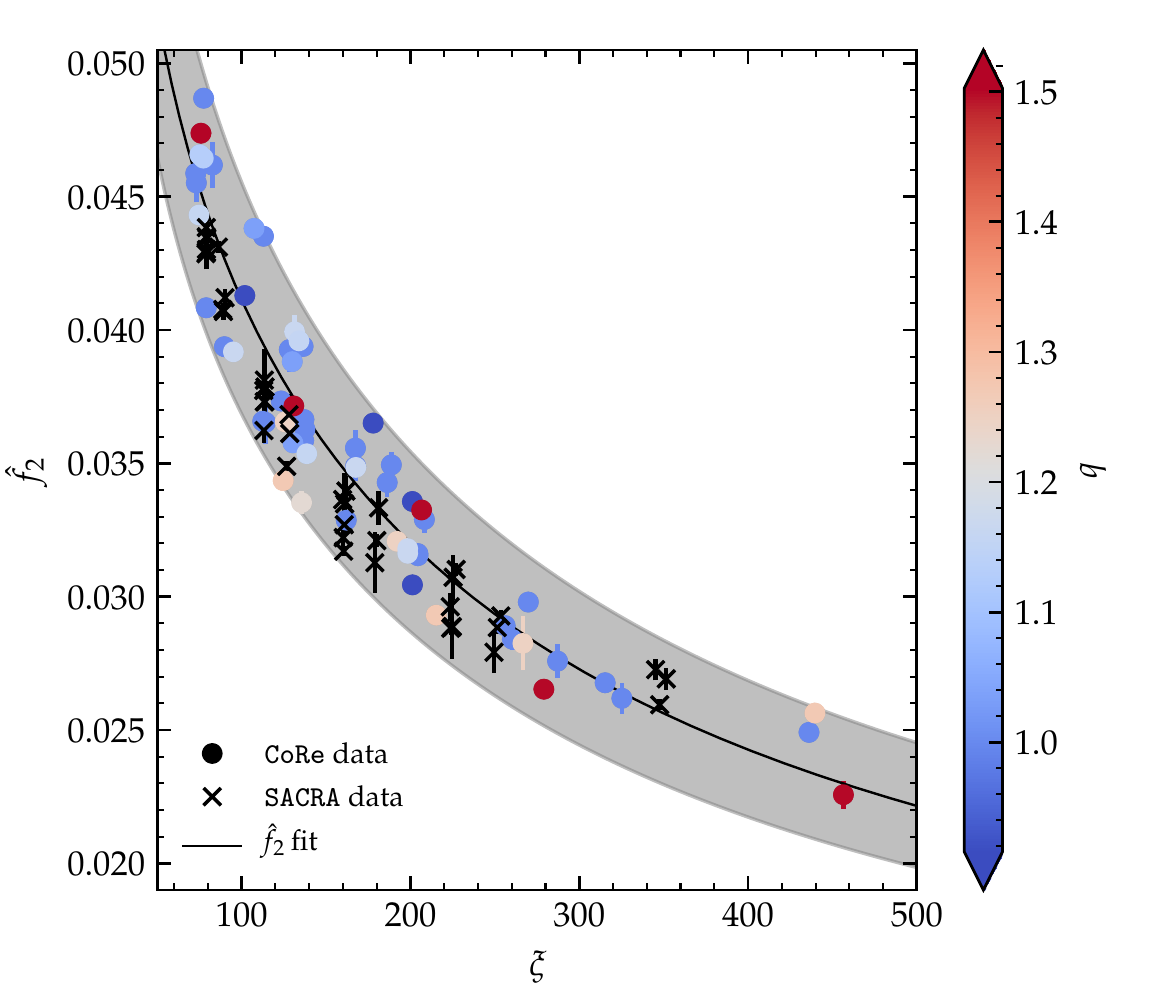}
  \caption{Phenomenological EOS-insensitive relation between the GW's main
    peak postmerger frequency and the (modified) tidal parameter
    $\xi(\kt2,\nu)$ (Eq.~\eqref{eq:xi}). 
    Both panels show the same data. The round markers correspond to
    the simulations of the {\tt CoRe} database. For those data the
    EOS variation is highlighted in colors in the left panel and the
    mass ratio variation in the right panel.
    The crosses correspond to the 
    {\tt SACRA} database
    that also refer to a large variation of EOS (although not highlighted
    in the graphics).
    Note that the frequency is the mass rescaled one in
    dimensionless units, $\hat{f}_{2}=Mf_{2}$.
    The fit is performed only on the {\tt CoRe} data and the gray band
    represents the 90\% confidence region.   
    Figure adapted from~\cite{Breschi:2019srl}.}
  \label{fig:fpm}
\end{figure}

The postmerger peak frequencies approximately correlate with the properties of the
binary and to properties of the nonrotating NS equilibria constructed
with the same EOS, e.g.
\cite{Bauswein:2011tp,Hotokezaka:2013iia,Bauswein:2014qla,Takami:2014zpa,Bernuzzi:2015rla,Lehner:2016wjg,Rezzolla:2016nxn,Kiuchi:2019kzt} 
(see also \cite{Bauswein:2019ybt} for a review).
EOS-insensitive phenomenological descriptions of the postmerger GW  
are thus possible.
As an example, Fig.~\ref{fig:fpm} shows a representation of the peak
postmerger frequency in terms of $\kt2$ \cite{Bernuzzi:2015rla}.
The basic idea behind this model is that the angular
momentum available at merger determines the rotation $\Omega$ of
the bulk mass, and that the GWs are efficiently radiated in short time
at this frequency. 
We stress again that the postmerger waveform is not formed by a set of
discrete frequencies but rather the frequency evolves continuously in
a nontrivial way, increasing (in a time-averaged way) in time as the remnant
becomes more compact. 
EOS-insensitive relations are the base to construct 
simple analytical representations of the postmerger GW \cite{Hotokezaka:2013iia,Bauswein:2015vxa,Bose:2017jvk,Tsang:2019esi,Breschi:2019srl}.
The use of these models to constrain matter at extreme densitites using kiloHertz GWs is explored in various works, e.g. \cite{Bauswein:2014qla,Steiner:2015aea,Torres-Rivas:2018svp,Breschi:2019srl}. 

The GW luminosity depends strongly on the merger remnant, as
illustrated by Fig.~\ref{fig:mrgwf}. For prompt collapse mergers the GW luminosity
peaks are the largest and happen shortly after the moment of
merger. Short-lived remnants have multiple 
peaks of comparable luminosity on a time scale of a few
milliseconds postmerger \cite{Zappa:2017xba}. These luminosity peaks correlate with
the oscillations of the instantaneous GW frequency (see middle panel
of Fig.~\ref{fig:mrgwf}) and correspond to the bounces of the NS cores.
Long-lived (and stable) NS remnants are qualitatively
similar to the short-lived ones but the GW emission is less intense due to 
the smaller compactness.  

A main difference with respect to binary black holes is that the most
luminous mergers do not correspond, in general, to those that radiate
the largest amount of energy. 
The largest GW energies per unit mass are
radiated by short-lived remnants over typical timescales of a few tens of 
milliseconds after the moment of merger \cite{Bernuzzi:2015opx}. This
is because a bar-deformed remnant NS close to gravitational collapse
is a very efficient emitter of GWs. The analysis of the energetics
from the simulations 
indicates that about two times the energy emitted during the inspiral and
merger can be emitted during the postmerger phase. This
is shown for a representative BNS in Fig.~\ref{fig:quasiu}: the binding
energy at the moment of merger is $-E_b\sim0.07$, while after the
postmerger transient is $-E_b\sim0.12-0.16$.
While the merger energy and peak luminosity tightly correlate with $\kt2$,
the total GW energy emitted by the remnant has a more complex behaviour. 
An absolute upper limit to the GW energy estimated by about one hundred simulations of the \texttt{CoRe} collaboration is \cite{Zappa:2017xba}
\be
E_{\rm GW} \lesssim 0.126\, \left(\frac{M}{2.8\Msun}\right)~\Msun {\rm
  c}^2 \ .
\ee

\begin{figure}
  \includegraphics[width=.49\textwidth]{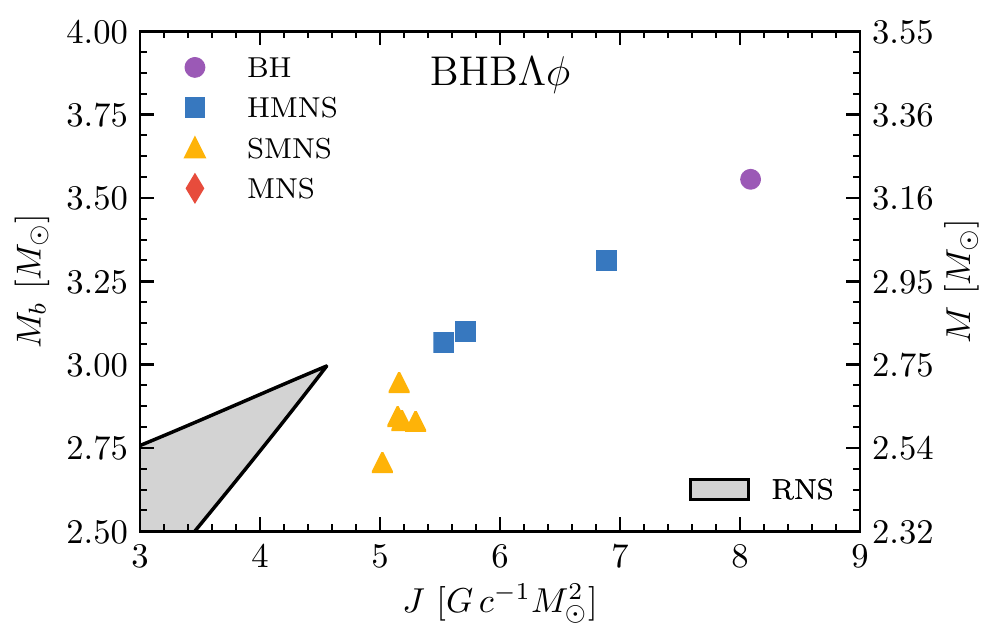}
  \includegraphics[width=.49\textwidth]{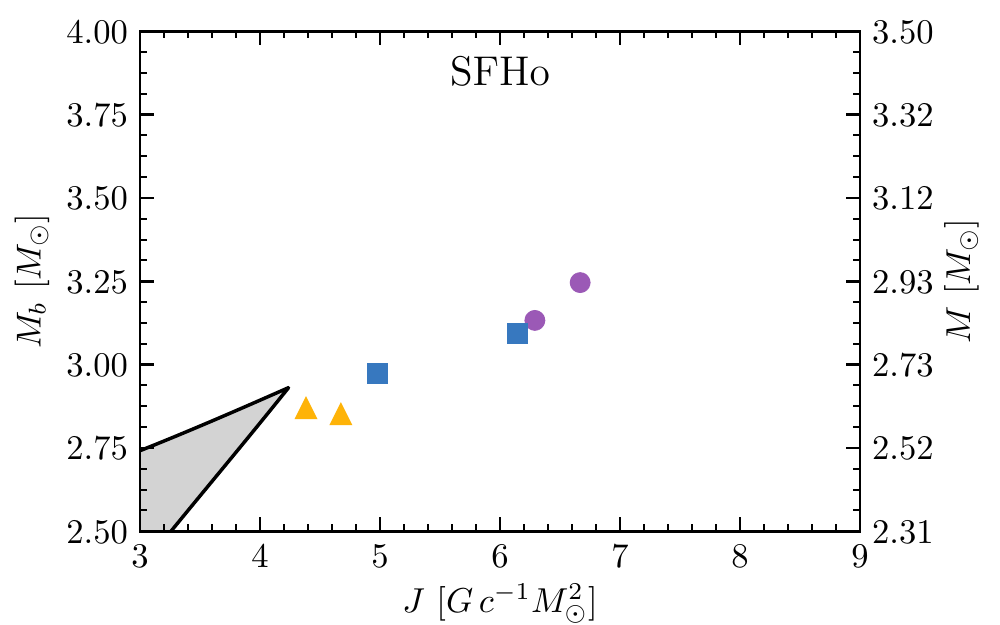}\\
  \includegraphics[width=.49\textwidth]{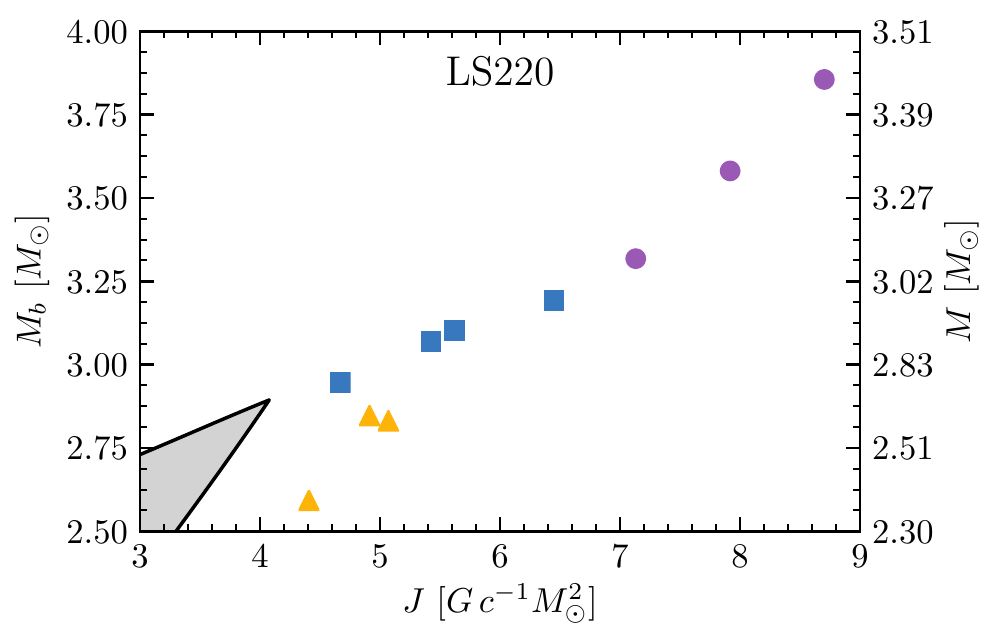}
  \includegraphics[width=.49\textwidth]{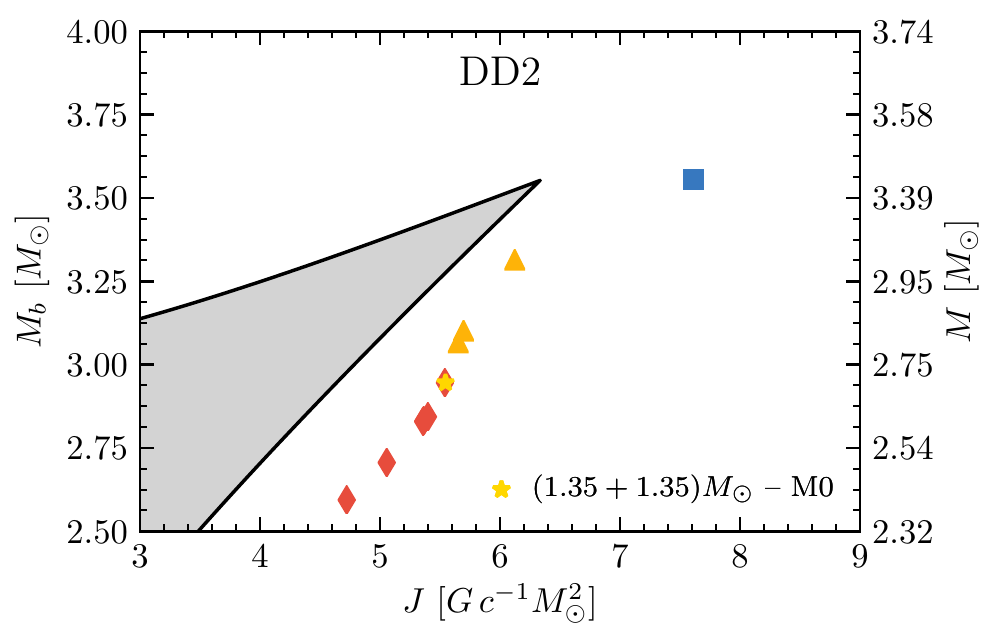}
  \caption{Diagrams of the rest-mass vs. angular momentum for
    representative merger remnants after the GW-dominated phase.
    The markers indicate remnants from fiducial mergers
    $M\sim2.7-2.8\Msun$, $q\sim1$ and microphysical EOS. The gray region is
    the stability region of rigidly rotating equilibria constructed assuming zero
    temperature and neutrino-less beta equilibrium
    \cite{Stergioulas:1994ea}. For a fixed $J$ 
    lower and upper boundaries of the shaded areas are set by the mass
    shedding and maximum mass limit, respectively. The tip of the shaded
    region marks the maximum baryonic mass configuration supported by each
    EOS in the case of rigid rotation.
    Figures from~\cite{Radice:2018xqa}.}
  \label{fig:MJ}
\end{figure}

\paragraph{Secular (Viscosity-driven) phase.}

As the GW emission of energy and angular momentum backreacts on the
fluid, it quickly damps nonaxisymmetric modes in the remnant that
evolves towards   
axisymmetry. The GW-emission timescale 
estimated at the end of the dynamical phase is \cite{Radice:2018xqa}
\be\label{eq:GWtau}
\tau_{\rm GW} = \dfrac{J}{\dot{J}_{\rm GW}}\gtrsim 0.5\, {\rm s} \ .
\ee
At this point the dynamics become dominated by viscous and cooling processes.

The NS remnants that did not collapse to BHs have angular momenta
significantly exceeding the mass-shedding limit for rigidly rotating NSs \cite{Radice:2018xqa}. 
Figure~\ref{fig:MJ} shows a diagram of the baryon mass $M_b$ and the angular
momentum $J$ of the remnant for a sample of remnants, and it compares them to the
rigidly rotating zero-temperature and beta-equilibrated isolated NS
equilibria (gray shaded region). The GW-driven phase in mergers'
remnants always ends on the right of the shaded areas; 
these remnants could be called ``super-Keplerian''.
Moreover, long-lived remnants have gravitational masses ${\sim}0.08\, \Msun$
larger than the corresponding equilibrium models having the same
baryonic (rest) mass, but zero temperature \cite{Radice:2018xqa}.
A key open question for future simulations is the evolution of
these systems on timescales of hundreds of milliseconds to seconds postmerger. 

The remnant evolution is determined by magnetohydrodynamics processes
and neutrino cooling and heating that affect the NS rotation and its temperature.
On the one hand, finite temperature and finite neutrino chemical
potential contribute to ${\sim}10\%$ increase of the pressure in the
NS core \cite{Kaplan:2013wra,Perego:2019adq}. Note that this is not sufficient to
significantly alter the maximum nonrotating mass due to the degeneracy
of matter above $\rho_0$.  
On the other hand, thermal support inflates the regions with
subnuclear densities increasing the NS radius. 
For characteristic  temperatures, the radius 
of a fiducial NS of mass $1.4\Msun$ increases by about $20-40\%$ (depending on
the EOS) compared to the zero-temperature nonrotating case.

Rotational support also increases the maximum NS mass. For example, in
the limiting case of rigid rotation at the Keplerian limit, the
maximum NS mass is increased by ${\sim}20\%$ with respect to a nonrotating
NS. 
Since this affects the whole star, the NS radius is typically increased by ${\sim}
40\%$, but at the same time the central density is decreased by a
similar amount if one compares nonrotating and Keplerian NSs of
identical masses. 
Interestingly, at temperatures reached in
merger remnants, the maximum mass for a stable rigidly-rotating
``hot'' NS remnant is actually smaller than that for cold
equilibria \cite{Kaplan:2013wra}. 
Rigidly-rotating NSs with temperature profiles similar to
those found in simulations can support ${\sim}0.1 M_\odot$ less baryonic
mass than cold configurations. 
While it is unlikely that finite temperature and composition effects
can stabilize a merger remnant against gravitational collapse, larger
radii imply that the mass shedding limit is reached at lower angular
frequencies. 
Hence, a NS remnant classified SMNS according to the
cold EOS could actually collapse to a BH. Alternatively, it might be 
possible to form stable NS remnants with baryonic masses and
thermodynamics profiles for which there is no rigidly-rotating
equilibrium. 

Magnetic fields also introduce additional pressure and can increase
the maximum mass and the maximum velocity of a rigidly rotating isolated
NS. However, the changes in maximum mass are moderate and up to $15-30\%$ for extreme
values of the magnetic field $B\sim10^{18}$~G \cite{Bocquet:1995je}. 
In merger remnants, magnetohydrodynamics instabilities and
magnetic-field amplifications can lead to global-scale magnetic
effects and angular momentum redistribution \cite{Duez:2006qe,Anderson:2008zp,Giacomazzo:2010bx,Kiuchi:2017zzg,Ciolfi:2019fie,Ruiz:2020via}. 
These instabilities operate on length scales of meters to centimeters,
and it is presently impossible to perform fully-resolved, global
merger simulations with realistic initial conditions. 
High-resolution simulations of mergers with magnetar-strength magnetic
fields showed that the Kelvin-Helmholtz instability at merger
could amplify the magnetic-field energy to up to 1\% of the thermal
energy \cite{Kiuchi:2015sga}. Moreover, if turbulent stresses are
modeled by an effective $\alpha$-viscosity, these simulations estimate 
$\alpha\simeq 0.01{-}0.02$ at $\rho \lesssim 10^{13}~{\rm g~cm^{-3}}$
(disc's densities) and $\alpha \sim 10^{-4}{-}10^{-3}$ at higher densities
\cite{Kiuchi:2017zzg}.  
Assuming the $\alpha-$viscosity model~\cite{Shakura:1972te}, the 
angular momentum redistribution in the remnant
happens on a timescale \cite{Hotokezaka:2013iia}:
\be
\tau_{\rm visc} \simeq \alpha^{-1}~R_{\rm rem}^2~\Omega_{\rm rem}~c_s^{-2} 
\simeq 0.56~{\rm s} 
\left( \frac{\alpha}{0.001} \right)^{-1}
\left( \frac{R_{\rm rem}}{15{\rm km}} \right)^2
\left( \frac{\Omega_{\rm rem}}{10^4{\rm kHz}} \right)
\left( \frac{c_s}{0.2c} \right)^{-2} \, ,
\label{eq:trem}
\ee
where $\Omega_{\rm rem}$ and $c_s$ are the remnant angular velocity
and typical sound speed, respectively. 
Simulations including a prescription for treating viscosity in GR find that the
remnant becomes more quickly axisymmetric, possibly reducing the
postmerger GW emission \cite{Radice:2017zta,Shibata:2017xht}. 
In particular, the turbulence induced by the magnetic field favours angular momentum
redistribution and accelerates the collapse or significantly affects the
remnant lifetime \cite{Hotokezaka:2013iia,Radice:2017zta}.
The magnitudes of these effects depends on the particular value
assumed for the $\alpha$-viscosity subgrid model.
For example, the use of a turbulence model calibrated to the 
high-resolution MHD runs of \cite{Kiuchi:2017zzg},
leads to significant changes to the subdominant features of the GW
spectrum and to the ejecta \cite{Radice:2020ids}. However, neutrino
effects on the ejecta are comparatively more relevant than 
magnetohydrodynamical turbulence.

The angular momentum redistribution in the remnant leads to characteristic rotational
profiles with a local minimum at the center \cite{Shibata:2006nm,Kastaun:2014fna,Endrizzi:2016kkf,Kastaun:2016yaf,Hanauske:2016gia,Ciolfi:2017uak}.
Since hydrodynamical and viscous effects counteract the gravitational instability of the core, the remnant's core is expected to spin up and to reduce its compactness \cite{Radice:2018pdn,Radice:2017zta}. 
This suggests that a super-Keplerian remnant evolving towards equilibrium must shed
excess angular momentum. 
Because the angular momentum losses cannot be GW-driven (Eq.~\eqref{eq:GWtau})
they must be driven by viscous effects on
timescales of $\tau_{\rm visc}$ and other electromagnetic processes that can extract the rotational 
energy of the remnant, e.g.~\cite{Metzger:2006mw,Siegel:2015swa}.
These processes can very efficiently generate large
outflows because the mass shedding limit moves to lower angular
momenta with
decreasing rest-mass $M_b$
\cite{Fujibayashi:2017xsz,Radice:2018pdn}.

Simulating 
the timescales $\tau_{\rm visc}$ is challenging for ab-initio numerical
simulations, so such a regime is currently  
explored in simplified setups (Newtonian gravity, axisymmetry, ad-hoc
initial conditions, etc., see Sec.~\ref{sec:disc}).
Together with viscous processes, neutrino interactions are the other
key process for the remnant evolution. The main effect is cooling, that
operates on timescales up to $\tau_{\rm cool}\sim2-3$~s \cite{Eichler:1989ve,Ruffert:1996by,Rosswog:2003rv,Sekiguchi:2011zd}.  
In addition, the excess of gravitational binding energy in the remnant found in NR
simulations is likely radiated in the form of neutrinos.  
These conditions are analogous to those found in newly born NSs in
core-collapse supernovae (e.g. \cite{Burrows:1981zz,Burrows:1986me,Pons:1998mm,Fischer:2009af,Roberts:2016mwj}).  

A possible outcome of the viscous evolution of a long-lived remnant is
a rotating NS close to the mass shedding limit with spin periods $P_0
\lesssim 1\, {\rm ms}$.  Comparing possible evolution scenarios to
equilibrium sequences, it is possible to estimate \cite{Radice:2018xqa}
\begin{equation}
\label{eq:spin}
  P_0 = \left[a \left(\frac{M_b}{1\, M_\odot} - 2.5\right) + b\right]
  {\rm ms}\ ,
\end{equation}
with EOS-dependent coefficients $a\sim-(0.2{-}0.3)$ and $b\sim1$.
Note that the above estimate gives spin periods significantly smaller than those
typically inferred for the progenitors of SGRB with extended
emission in the context of the magnetar model, $P_0 \sim 10\, {\rm ms}$ 
\cite{Fan:2013cra,Gompertz:2013aka}.
Gravitational-wave losses could however continue past the viscously-driven
phase of the evolution and further spin down the remnant over a timescale of
many seconds to minutes \cite{Fan:2013cra,Gao:2015xle}. The GW
emission could be driven by secular instabilities in the remnant
\cite{Friedman:1975,Chandrasekhar:1992pr,Lai:1994ke,Cutler:2002nw,Stergioulas:2003yp,Corsi:2009jt,DallOsso:2014hpa,Doneva:2015jaa,Lasky:2015olc,Paschalidis:2015mla,East:2015vix,Radice:2016gym,Lehner:2016wjg,East:2016zvv}
(see also \cite{Stergioulas:2003yp} for a review), or by deformations due to a strong toroidal field \cite{Fan:2013cra}.
For example, the GW luminosity of the one-armed instability
during the first ${\sim}50\, {\rm ms}$ of the post-merger evolution is
${\sim}10^{51}\, {\rm erg}\, {\rm s}^{-1}$ and does not show strong
evidence for decay \cite{East:2015vix,Radice:2016gym}. If the one-armed instability
were to persist without damping, then it would remove all of the NS
remnant rotational energy, which is ${\sim}10^{53}\, {\rm erg}$
\cite{Margalit:2017dij}, over a timescale of ${\sim}100\, {\rm
s}$. This timescale is compatible with the spin-down timescale inferred
from the magnetar model \cite{Fan:2013cra}. This GW signal
from the one-armed instability could be detectable by LIGO-Virgo up
to a distance of ${\sim}100\, {\rm Mpc}$ for optimally oriented sources
\cite{Radice:2016gym}.

\section{Remnant discs}
\label{sec:disc}

\begin{figure}
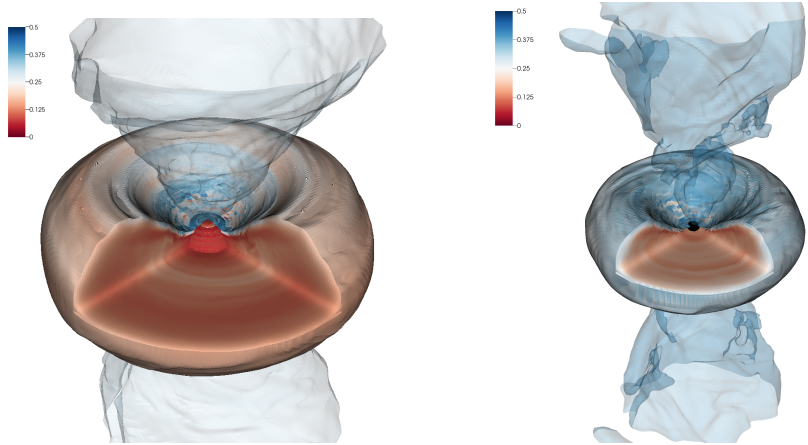

  \includegraphics[width=.49\textwidth]{fig06a.pdf}\qquad
  \includegraphics[width=.49\textwidth]{fig06b.pdf}
  \caption{Example of discs around NS (left) or BH (right) remnants. 
   The figure shows a 3D rendering of the electron fraction for
    equal-masses BNSs described by the DD2 (left) and SFHo (right).
    Both images have the same spatial scale and show the data in a box
    of size $750$~km. The electron fraction is used to color the
    $10^7$~$\gccm$ (semi-transparent) and the $10^{11}$~$\gccm$ density
    isosurfaces. The $10^{13}$~$\gccm$ isosurface is also shown for the
    DD2 model. The black surface in the SFHo  model denotes the
    approximate location of the black hole horizon.
    The discs are fairly neutron rich in their bulk,
    irrespective of the remnant type (massive NS or black hole). The
    accretion disc coronae are irradiated by neutrinos and are less
    neutron rich.  
    Figure from~\cite{Perego:2019adq}.}
\label{fig:disk3d}
\end{figure}

Following a common convention, the remnant disc is here defined as the baryon
material either 
outside the BH's apparent horizon or that with densities 
$\rho\lesssim10^{13}$~$\gccm$ around a NS remnant. The baryonic mass
of the disc is computed in simulations as volume integrals of the
conserved rest-mass density and it is referred to as $M_{\rm disc}$.
Remnant discs are geometrically thick discs with 
typical aspect ratio $H/R \sim 1/3$ and mass between
$0.001{-}0.2\ \Msun{}$. The structure and composition of the remnant discs can
significantly depend on the different formation mechanisms due to the
different binary properties.

In the case of comparable mass mergers, the
accretion disc is formed during and after the merger by the matter
expelled by tidal torques and by the collision interface. 
Because of the different temperatures in the tidal tail (cold) and
collisional interface (hot), the disc is initially highly
non-uniform.
As time evolves, the NS remnant continuously 
sheds mass and angular momentum into the disc with spiral density
waves as described in Sec.~\ref{sec:ns}, thus increasing the mass of the disc and
generating mass outflows 
\cite{Bernuzzi:2015opx,Radice:2018xqa} (see also Sec.~\ref{sec:ejecta}).  
The continued action of shocks and spiral waves increases the entropy
in the disc and eventually produces an approximately axisymmetric Keplerian disc
characterized by a temperature profile that changes smoothly from
$\sim10\,$MeV (for $ \rho \simeq 10^{13}\,$$\gccm$) down to
${\sim}0.1\,$MeV (for $ \rho \simeq 10^{4}\,$$\gccm$). The electron fraction
is reset by pair processes and 
the entropy per baryon varies between 
3 and several 10's of $k_{B}$/baryon \cite{Perego:2019adq}.
In general, BH formation significantly affects the disc properties, as
illustrated by Fig.~\ref{fig:disk3d}. If the central
object collapses to a BH, approximately half of the disk mass is
swallowed inside the apparent horizon within a dynamical timescale,
and the maximum density decreases to a few times $10^{12}$~$\gccm$. Discs around a BH remnant are in general more compact and
achieve higher temperatures and entropies ($\Delta s \simeq
2~{k_B/{\rm baryon}}$) than discs hosting a NS remnant. 

\begin{figure}
  \includegraphics[width=.9\textwidth]{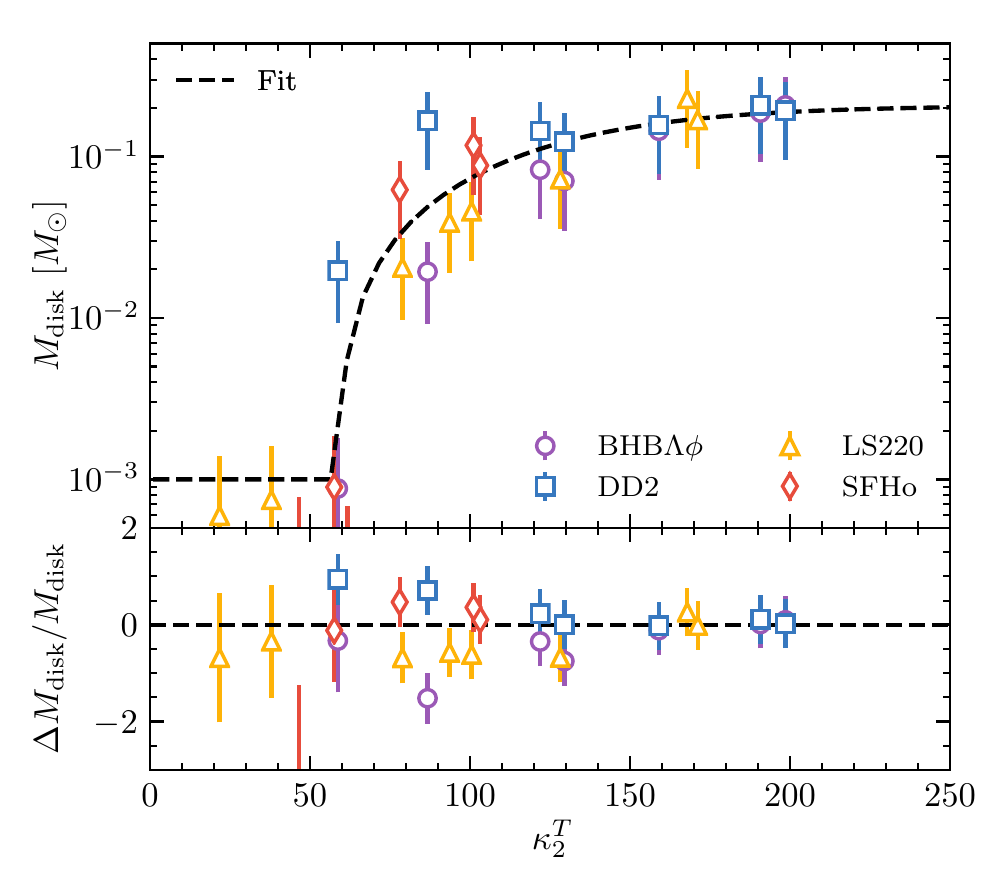}
  \includegraphics[width=.9\textwidth]{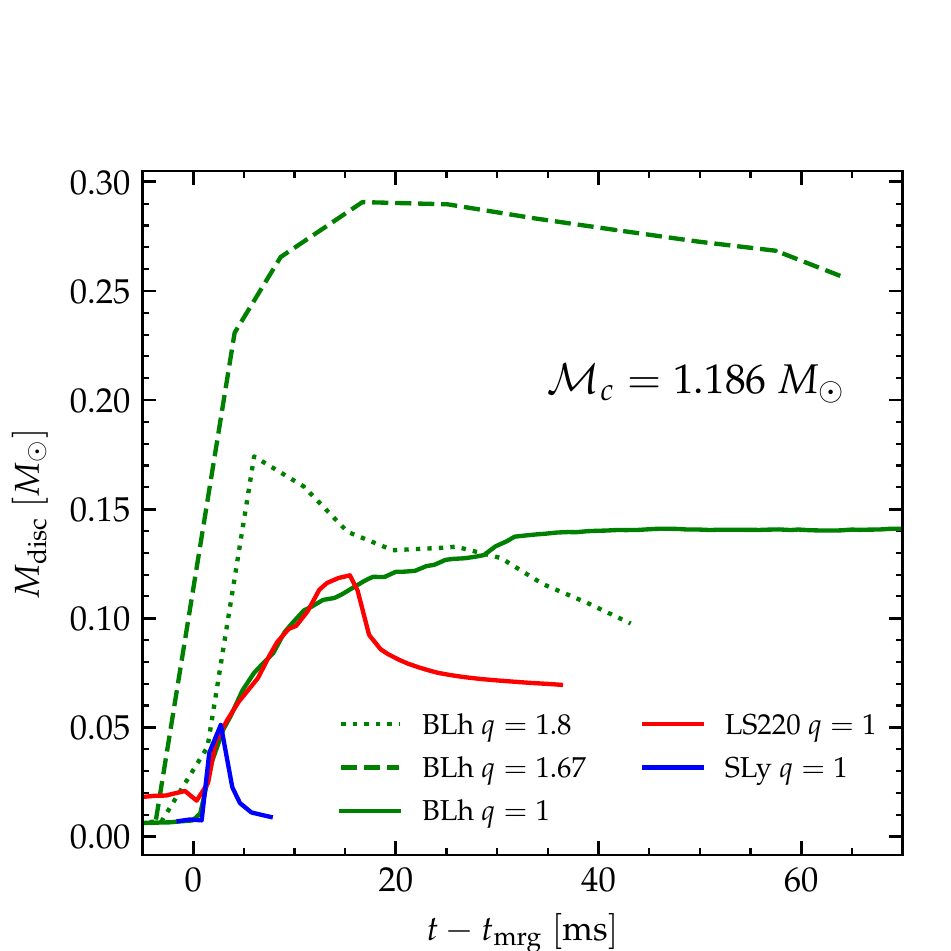}
  \caption{Disc masses as a function of the tidal parameter and
    disc mass evolution for representative cases.
    Top: The remnant disc mass of equal-mass mergers 
    correlates with $\kt2$, the latter measuring the binary
    compactness. Small values $\kt2\lesssim\kappa_{\rm pc}$ correspond
    to prompt collapse mergers for which a disc with negligible mass forms.
    Figure adapted from~\cite{Radice:2018pdn}. The bottom panel shows 
    the relative difference between the data and the fit.
    Bottom: Disc formation and early evolution for mergers with chirp
    mass $1.186\Msun$. The $q=1$ SLy and LS220 are short-lived
    remnants collapsing to BH within 2 and 18 milliseconds
    respectively. The BLh $q=1,1.67$ are long-lived remnants, while
    the $q=1.8$ is an accretion induced prompt collapse.
    Figure adapted from~\cite{Bernuzzi:2020txg}.}
\label{fig:disk}
\end{figure}

Disc masses at formation are shown in the top panel of Fig.~\ref{fig:disk}
as a function of the tidal parameter $\kt2$. Again, the choice of
$\kt2$ for this plot is for correlating the disc with a measure of the
binary compactness [Note however that the parameter is not a good
  choice for cases dominated by tidal distruption~\footnote{An
    extreme case is for example the disc mass in 
    black-hole--neutron star binaries, that does not show strong
    correlation  with $\kt2$,
    \cite{Foucart:2018rjc,Zappa:2019ntl}. See discussion in \cite{Breschi:2019srl}.}]. 
The figure highlights that for $q\sim1$ prompt collapse mergers do not form
massive discs (Cf. Eq.~\eqref{eq:kappapc}), because the mechanism
primarily responsible for the formation of the disc shuts off
immediately in these cases,
e.g. \cite{Shibata:2006nm,Kiuchi:2009jt,Radice:2017lry}. 
Short-lived and long-lived remnants have instead discs with initial masses ${\sim}0.2\,\Msun{}$.
Mergers of BNSs with mass ratios up to $q\sim1.3-1.4$ however produce more
massive discs than $q=1$ because of the larger centrifugal support and
a partial tidal disruption of the companion NS \cite{Shibata:2003ga,Shibata:2006nm,Kiuchi:2009jt,Rezzolla:2010fd,Dietrich:2016hky}.

In high mass-ratio mergers with $q\gtrsim1.5$ the companion NS is tidally
disrupted and the disc is mainly formed by the tidal tail \cite{Bernuzzi:2020txg}.
The latter is launched prior to merger and massive accretion discs are possible even if prompt 
BH formation occurs \cite{Kiuchi:2019lls,Bernuzzi:2020txg}.
The angular momentum of these discs can be ${\sim}60\%$ larger than
that of discs around BHs resulting from equal-mass mergers. 
Moreover, due to the absence of strong compression and shocks, the discs formed
in high mass-ratio mergers are initially colder and more neutron rich
than those of comparable-mass mergers having the same chirp mass.

Examples of disc mass evolutions at early times from formation for
different remnants are shown in the right panel of
Fig.~\ref{fig:disk}. The figure clearly shows the rapid accretion in
case of equal-masses ($q\sim1$) mergers and BH formation. Discs around
NS remnants instead can also increase their mass over time as the
remnant's spiral waves propagate outwards. The accretion of discs
around BHs formed in high-mass-ratio mergers is instead slower 
due to the larger disc's extension and angular momentum.
Here, however, accretion is further driven by the fallback of the
tidal tail that perturbs the disc inwards \cite{Bernuzzi:2020txg}.

The long-term evolution of these discs is key for electromagnetic
emission, and studies in this direction are becoming more complete
and detailed
\cite{Fernandez:2013tya,Metzger:2014ila,Just:2014fka,Fernandez:2015use,Fujibayashi:2017puw,Siegel:2014ita,Fujibayashi:2017xsz,Perego:2017fho,Fernandez:2018kax,Miller:2019dpt}.
However, the challenges related to
the simulation of multiples scales and multiple physical processes have, so far, 
required the adoption of some simplifications.
All of the published simulations either adopted somewhat artificial
initial conditions (not derived from merger simulations), neglected
important physical effects such as neutrino emission and absorption, 
assumed axisymmetry, or did not follow the evolution for
sufficiently long times.  
Crucial questions are related to the development of ordered large-scale
magnetic fields formed by dynamo processes and the interplay with
neutrino interactions.
Large-scale magnetic fields can power relativistic jets
\cite{Blandford:1977ds,VanPutten:1999vda,Bucciantini:2011kx,Ruiz:2016rai,Mosta:2020hlh} and drive
mildly relativistic outflows
\cite{Siegel:2014ita,Metzger:2018uni,Fernandez:2018kax}. 
Neutrinos emitted from the hottest and densest part of the remnant
irradiate the low density part of the disk (and the expanding
wind) 
thus increasing substantially the electron fraction in the
material \cite{Perego:2017fho}. The larger effects are for hotter
remnants and along the 
polar regions, where neutrino fluxes are more intense due to the lower
optical depths \cite{Wanajo:2014wha,Perego:2014fma,Goriely:2015fqa,Radice:2016dwd,Sekiguchi:2016bjd,Perego:2017fho,Martin:2017dhc,Radice:2018ozg,Endrizzi:2020lwl}.
The combined effect of magnetohydrodynamics and neutrino processes is
likely to play an important role in the dynamics and should be further
explored by future simulations \cite{Guilet:2016sqd,Siegel:2017jug,Mosta:2020hlh}.

\section{Mass ejecta}
\label{sec:ejecta}

Since the ejection of neutron rich material happens at different stages of
the merger dynamics, mass ejecta have multiple components with
different properties, geometries and composition
\cite{Shibata:2019wef,Rosswog:2017sdn}.

\paragraph{Dynamical ejecta.}
Dynamical mass ejecta are launched during 
the merger process. A fraction of the material is launched 
by tidal torques around the moment of merger
\cite{Rosswog:1998hy,Radice:2016dwd,Dietrich:2016hky}; 
another fraction of matter is unbound from shocks generated after the
moment of merger when the cores bounce \cite{Hotokezaka:2012ze,Bauswein:2013yna,Sekiguchi:2016bjd,Radice:2018pdn}. 
General-relativistic merger simulations indicate that the mass of the
dynamical ejecta ranges from $10^{-4}\ \Msun$ to 
$10^{-2}\ \Msun$ and that it has characteristic velocities of
$0.1{-}0.3$c \cite{Hotokezaka:2012ze,Bauswein:2013yna,Sekiguchi:2016bjd,Radice:2018pdn}. 
The tidal ejecta are neutron rich $Y_e \sim 0.1$ and cold, while
the shocked ejecta are reprocessed to higher $Y_e$ by pair processes
and neutrino irradiation from the 
NS remnant. The electron fraction in shocked ejecta can span a wide
range of values, $Y_e\sim0.1-0.4$, with the largest $Y_e$ obtained at
high latitudes.
If large-scale magnetic fields are present at the moment of merger,
they could additionally boost the dynamical (shocked) ejecta with a
viscous component \cite{Radice:2018ghv}.

\begin{figure}
  \includegraphics[width=.9\textwidth]{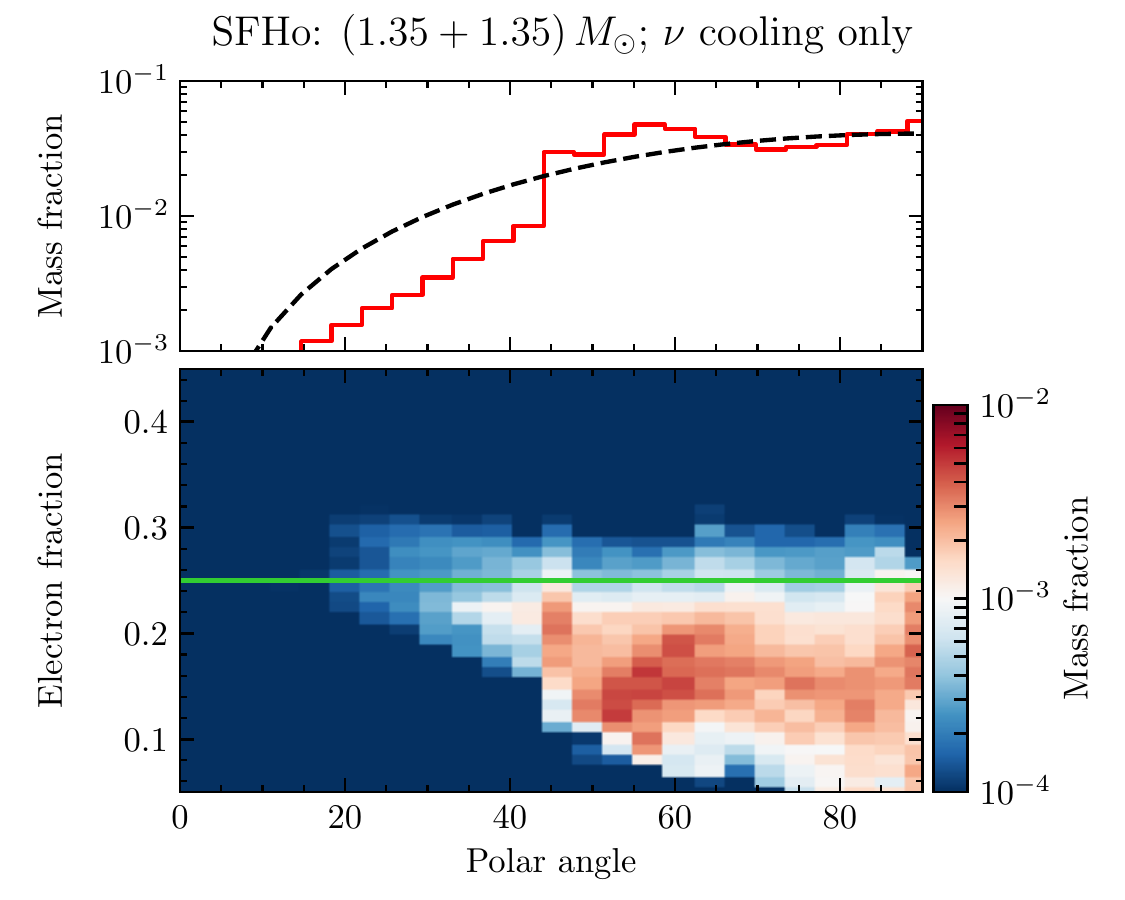}
  \includegraphics[width=.9\textwidth]{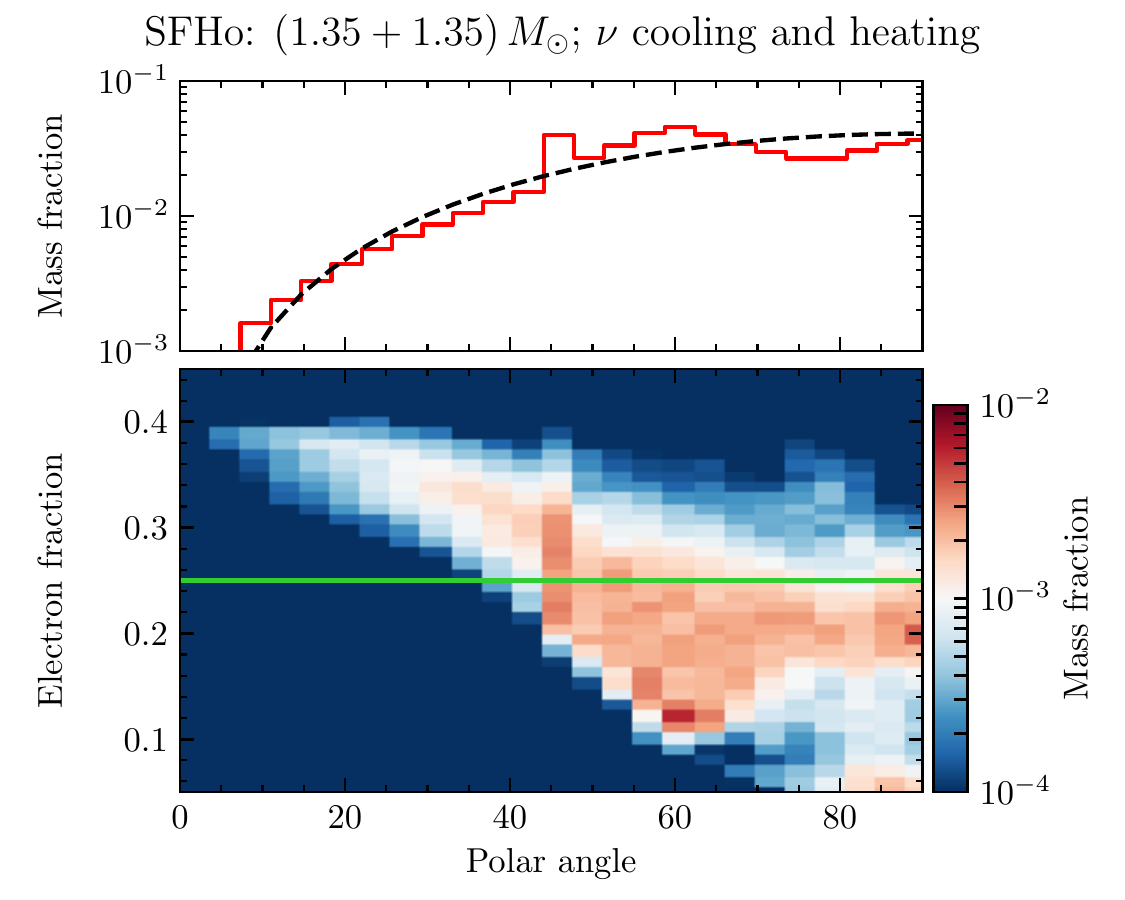}
  \caption{Dynamical ejecta mass and composition as a function of the
    polar angle $\theta$. The orbital plane is $\theta=90^\circ$.
    The top panel refers to a simulation with neutrino leakage only,
    the bottom panel to a simulation with neutrino leakage and the M0
    transport scheme for free-streaming neutrinos. The dashed black
    line refers to a model distribution $Y_e(\theta)\sim\sin^2\theta$.
    Figure adapted from~\cite{Perego:2017wtu}.}
\label{fig:dyn_ej_comp}
\end{figure}

For comparable-mass mergers, NR simulations
indicate that the shocked component is typically a factor ten more
massive than the tidal component.
This is in contrast to early works that employed Newtonian gravity and in which
the tidal component dominated the ejecta due to the weaker gravity and
stiffer EOS employed in those simulations
\cite{Ruffert:1996by,Rosswog:1998hy,Rosswog:2001fh,Rosswog:2003rv,Rosswog:2003tn,Rosswog:2003ts,Oechslin:2006uk,Rosswog:2013kqa,Korobkin:2012uy}.
A sample of about 130 NR simulations using microphysics EOS and approximate neutrino transport
indicate that ejecta masses do not strongly correlate in a simple way
with the properties of
the binary \cite{Radice:2018pdn}. The average mass is
${\sim} 2\times10^{-3}\ \Msun$ \cite{Radice:2018pdn}, 
the mass-averaged speed is about 
$\langle v_{\rm dyn}\rangle\sim0.18$c (although some part of the
ejecta can reach high-speeds up to $\sim0.8$c
\cite{Bauswein:2013yna,Radice:2018pdn}), and the mass-averaged electron fraction is $\langle Y^{{\rm  dyn}}_e\rangle\sim0.17$.
Neutrino absoprtion has a significant effect on the composition of
dynamical ejecta, and some radiation transport scheme that includes
neutrino absoprtion must be considered in the simulations for
a realistic estimate of $Y_e$ \cite{Wanajo:2014wha,Perego:2017wtu}.
The dynamical ejecta properties vary with the polar angle \cite{Perego:2017wtu}.
The mass is launched about the orbital plane with a r.m.s. of
${\sim}35^\circ$; the highest velocities and electron fraction are 
obtained at high latitudes where the medium densities are lower and the neutrino
fluxes are more intense. In particular, the electron fraction has a
profile well approximated by $Y_e\sim\sin^2\theta$, where $\theta$ is
the polar angle with the axis normal to the orbital plane.
As an example, Fig.~\ref{fig:dyn_ej_comp} shows the distribution of
mass and $Y_e$ in the polar angle, for two simulations that differ
only in the neutrino transport scheme employed. If a leakage scheme is
employed, thus only neutrino cooling is simulated, then the ejecta have
no material with $Y_e>0.25$. If also neutrino absorption is simulated,
then the $Y_e$ distribution extends to $Y_e\sim0.4$ in the region $\theta<60^\circ$.

\begin{figure}
  \includegraphics[width=.9\textwidth]{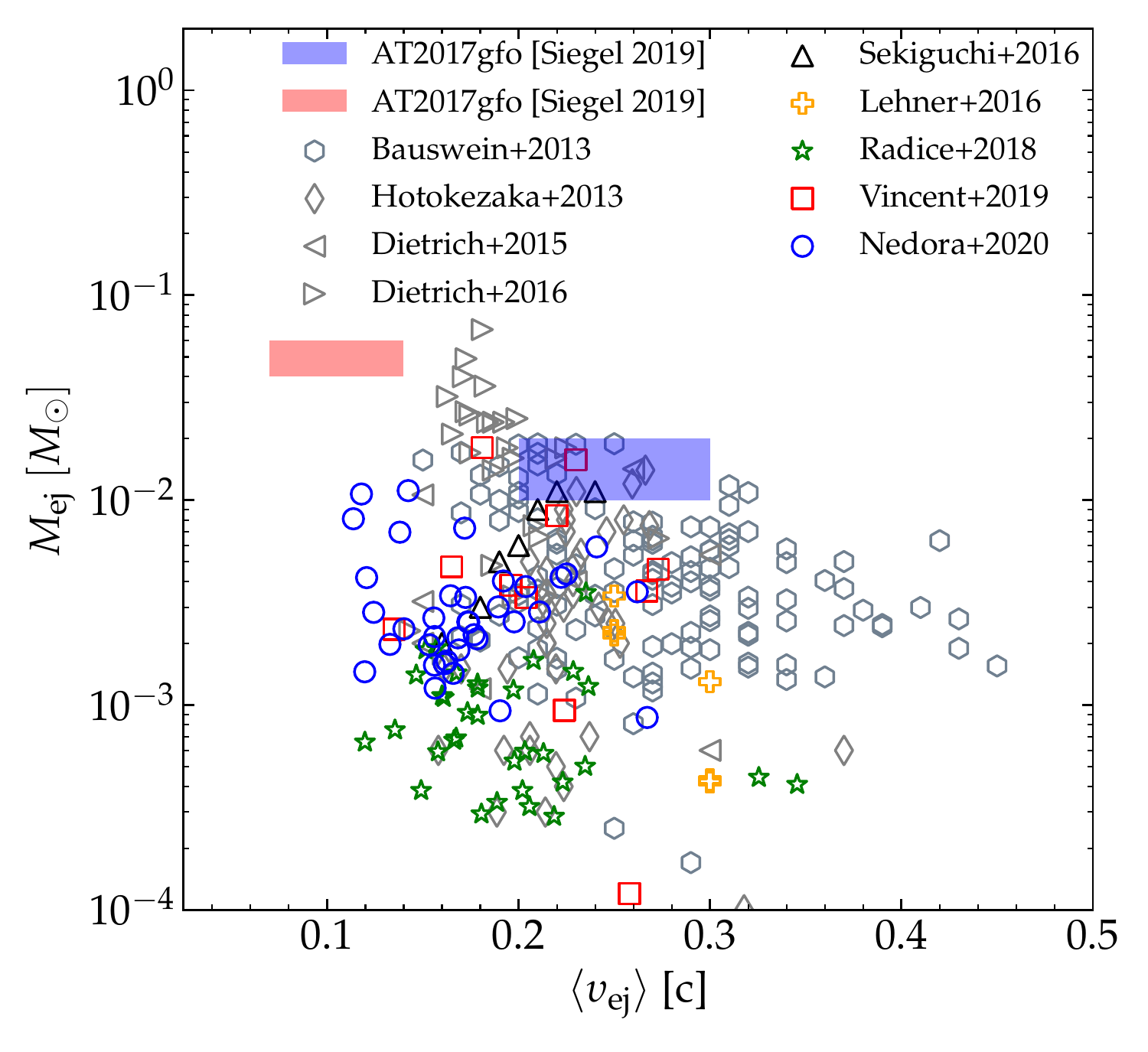}
  \includegraphics[width=.49\textwidth]{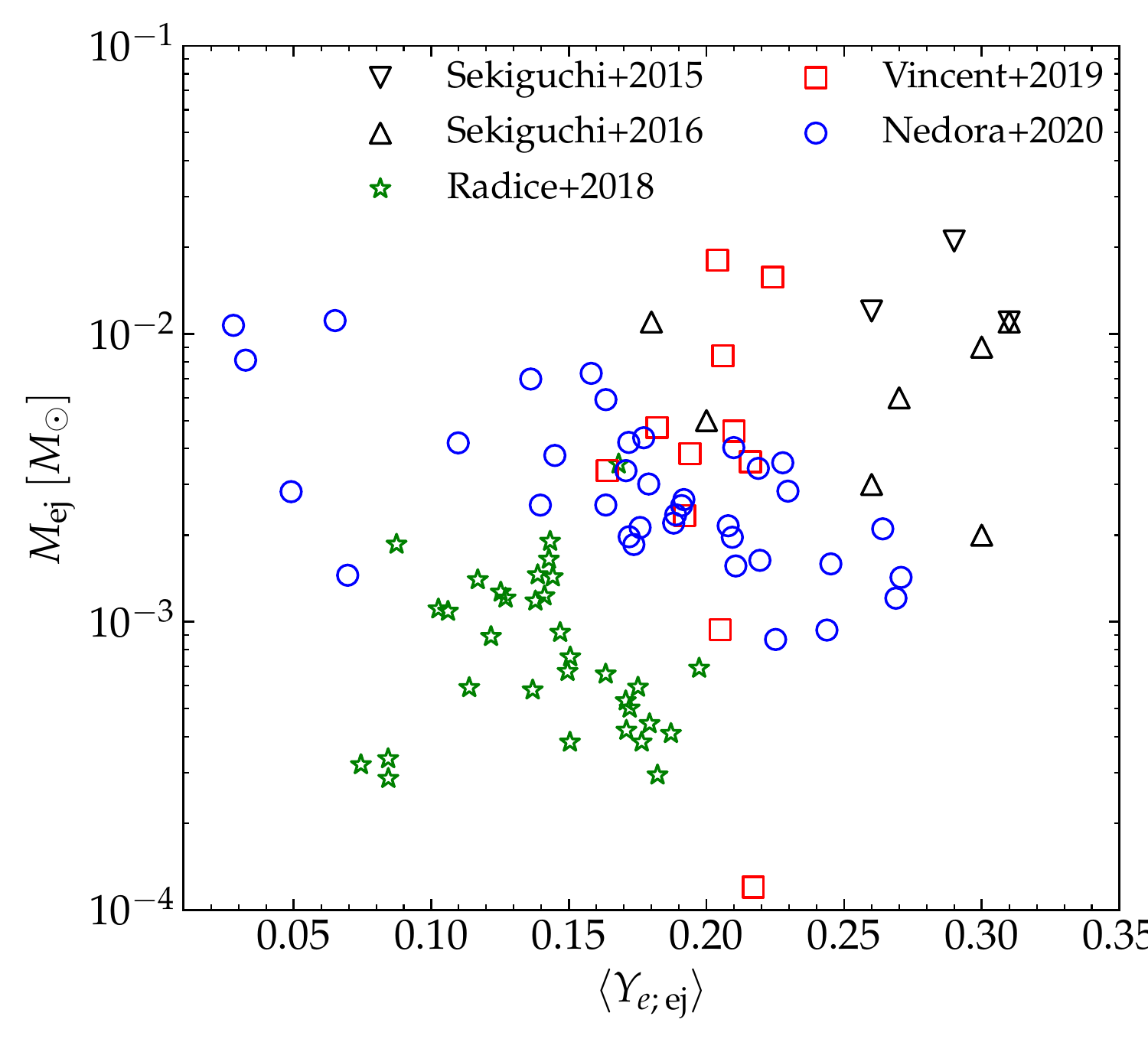}
  \includegraphics[width=.49\textwidth]{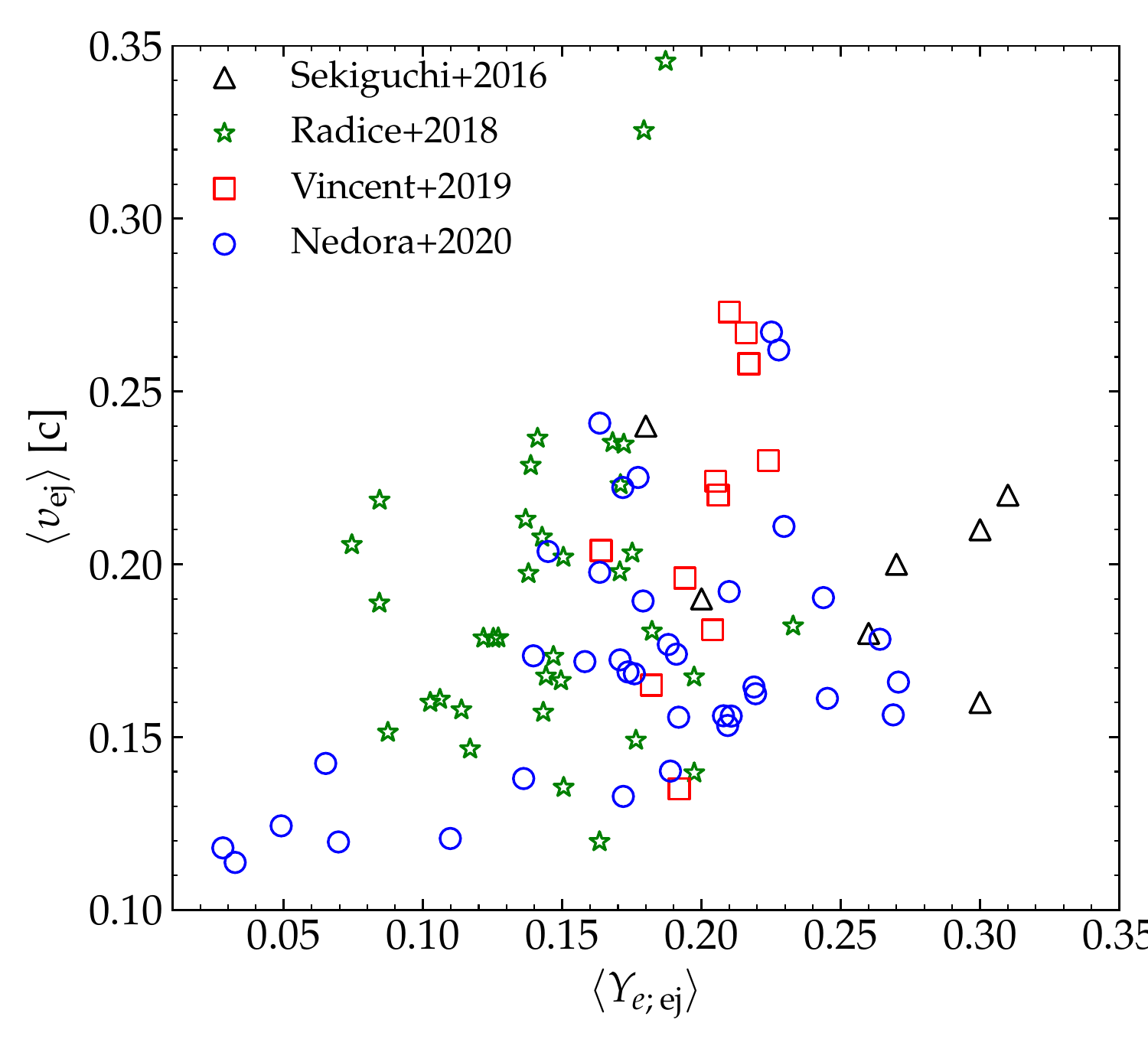}
  \caption{Summary of the dynamical ejecta properties (mass, mass-averaged
    velocity and electron fraction) as found by simulations
    with different physics input, different NS masses in ${\sim}1.2-1.5\Msun$ and EOS. The datasets include: polytropic EOS
    data from \cite{Hotokezaka:2012ze,Dietrich:2015iva,Dietrich:2016lyp,Kiuchi:2019lls},
    microphysical EOS data with no neutrinos treatment from \cite{Bauswein:2013yna},
    microphysical EOS data with leakage scheme from
    \cite{Sekiguchi:2015dma,Lehner:2016lxy,Radice:2018pdn}, 
    microphysical EOS data with M1 or leakage+M1 scheme from
    \cite{Sekiguchi:2016bjd,Vincent:2019kor}, 
    microphysical EOS data with leakage+M0
    \cite{Radice:2018pdn,Perego:2019adq,Nedora:2019jhl,Bernuzzi:2020txg,Nedora:2020}.
    The filled blue and red patches are the expected values of
    ejecta mass and velocity for blue and red components of the 
    kilonova AT2017gfo, compiled by \cite{Siegel:2019mlp} and based on
    \cite{Villar:2017wcc}. Figure courtesy of V.Nedora.} 
\label{fig:dyn_ej_summary}
\end{figure}

The dynamical ejecta data obtained by different groups 
with independent codes, for similar binaries and input physics are broadly consistent within a
factor of two
\cite{Radice:2018pdn,Perego:2017wtu,Nedora:2019jhl,Endrizzi:2020lwl,Sekiguchi:2015dma,Palenzuela:2015dqa,Lehner:2016lxy,Shibata:2017xdx,Bovard:2017mvn,Vincent:2019kor}. Numerical
errors can account for the difference in some cases, but for the highest resolutions simulated
so far the numerical uncertainties are around 20-40\%
\cite{Shibata:2017xdx,Radice:2018pdn,Bernuzzi:2020txg}.
Figure~\ref{fig:dyn_ej_summary} collects the dynamical ejecta
properties for a representative sample obtained by various groups
using different physics assumptions. In particular, it includes: (i) the
piecewise-polytropic EOS runs of 
\cite{Hotokezaka:2012ze,Dietrich:2015iva,Dietrich:2016lyp,Kiuchi:2019lls},
in which temperature effects are approximated by a $\Gamma$-law EOS and
composition and weak effects are not simulated;
(ii) the microphysical EOS data of \cite{Bauswein:2013yna}
in which weak reactions are not simulated;
(iii) the microphysical EOS data of \cite{Sekiguchi:2015dma,Lehner:2016lxy,Radice:2018pdn}
in which a leakage scheme is employed for neutrino cooling;
(iv) the microphysical EOS data of \cite{Sekiguchi:2016bjd,Vincent:2019kor}
in which a leakage+M1 scheme and a M1 gray scheme respectively are
employed for the neutrino transport;
(v) the microphysical EOS data of in which a leakage+M0 scheme are
employed for the neutrino transport
\cite{Radice:2018pdn,Perego:2019adq,Nedora:2019jhl,Bernuzzi:2020txg,Nedora:2020}. 
The largest differences in the computations reported
in the literature are related to the use of different input physics.
Microphysics and neutrino absorption have a significant impact on the
dynamical ejecta properties
\cite{Wanajo:2014wha,Sekiguchi:2015dma,Perego:2017wtu,Vincent:2019kor},
as evident from Fig.~\ref{fig:dyn_ej_comp}. Microphysical EOS determine  
average velocities smaller than those computed using polytropic EOS,
and distributed up to $0.3$c.
The inclusion of neutrino absorption results in larger average ejecta
masses and electron fractions then those obtained with the leakage scheme.
Simulations with polytropic EOS or without neutrino leakage,  
e.g. \cite{Hotokezaka:2012ze,Bauswein:2013yna,Dietrich:2015iva}, give
ejecta masses up to factor five larger than those obtained with simulations
with microphysics and neutrino transport schemes, and in some cases
average velocities up to $\langle v_{\rm dyn}\rangle\sim0.3-0.4\,$c.
Phenomenological fitting formulas of dynamical ejecta properties in terms 
of the binary parameters are presented in
\cite{Dietrich:2016fpt,Radice:2018pdn,Kruger:2020gig}.
Note that the different fits are not fully consistent with each other;
they depend on the particular datasets employed and some trends appear in the residuals.
Simulations including microphysics and neutrino effects and spanning different chirp masses and mass ratios 
are required in order to quantify a clear dependence of the ejecta on the binary properties.

For highly asymmetric BNSs with $q\gtrsim1.67$ the dynamical ejecta is
instead dominated by the tidal component
\cite{Sekiguchi:2015dma,Lehner:2016lxy,Bernuzzi:2020txg}. 
Here the ejecta is distributed more narrowly about the orbital plane and over
a fraction of the azimuthal angle around its ejection angle with a
crescent shape.
Extreme mass asymmetry can boost the mass ejecta by up
to a factor four with respect to the equal mass cases (for a fixed
chirp mass). In this case, the average electron fraction reduces to
${\sim}0.11$, and the 
r.m.s. of the polar angle is ${\sim}5-15^\circ$ \cite{Bernuzzi:2020txg}.
This is similar to what is observed in black-hole--neutron-star binaries,
\cite{Kawaguchi:2016ana}.

In asymmetric mergers of rotating NSs with spin aligned to the orbital angular
momentum, the dynamical ejecta mass can increase due to the larger angular
momentum of fluid elements in the tidal tail \cite{Dietrich:2016lyp}.
However, for equal-mass mergers the ejecta mass can decrease for large
aligned spins \cite{Dietrich:2016lyp,Most:2019pac} because at the
moment of merger the system is more bounded (smaller $j_{\rm mrg}$ and more negative
$E_b^{\rm mrg}$ as aligned-spin-orbit interaction is repulsive) and
less material is unbound from the core shock.
Overall, spin effects are sub-dominant with respect to mass ratio
effects \cite{Dietrich:2016lyp}.

\paragraph{Secular ejecta.}
Another type of ejecta is the secular winds originating from the
remnant
\cite{Lee:2009uc,Dessart:2008zd,Fernandez:2013tya,Metzger:2014ila,Just:2014fka,Perego:2014fma,Fernandez:2015use,Siegel:2017nub,Fujibayashi:2017puw,Fernandez:2018kax,Radice:2018pdn}.    
Long-term Newtonian simulations of neutrino-cooled accretion
disks indicate that $10{-}40\%$ of the remnant
disk can unbind over a timescale of a few seconds. 
Since remnant discs in mergers have masses up to
${\sim}0.2\ \Msun$, winds are likely to constitute the
bulk of the ejecta (if present).
These secular ejecta can originate from different physical mechanisms. 

Neutrinos from the remnant and the disc drive a wind of material with
$Y_e\sim0.3$ and can unbind ${\lesssim}10^{-3}\ \Msun$
\cite{Dessart:2008zd,Just:2014fka,Perego:2014fma,Miller:2019dpt}. 
The neutrino wind originates on the disc edge, close to the
neutrinosphere, and above the remnant where baryon pollution is minimal.
Note that a precise prediction of properties of polar ejecta is presently beyond the
possibilities of neutrinos schemes employed in ab-initio NR
simulations \cite{Just:2014fka,Fujibayashi:2017puw,Foucart:2018gis,Miller:2019dpt}. 

Long-term NR simulations have shown that, if the merger outcome is a
NS remnant, the spiral density waves propagaing from the remant into
the disc can trigger a massive and fast wind \cite{Nedora:2019jhl}. These ejecta start 
after the moment of merger and operate on timescales longer than the
dynamical ejecta. Their origin is purely hydrodynamical but viscosity and neutrino transport influence the angular momentum transported by the spiral waves and their composition.
The spiral wind
can have a mass up to ${\sim}10^{-2}\Msun$ and velocities
$\lesssim0.2$~c. The ejected material has electron fraction mostly
distributed above $Y_e\sim0.25$ being partially reprocessed by hydrodynamic
shocks in the expanding arms. 

Angular momentum transport due to viscous processes causes the disc to
spread outwards. Once the accretion rate drops below a critical threshold,
neutrino cooling becomes ineffective and the disc thermally expands
\cite{Beloborodov:2008nx,Lee:2009uc,Fernandez:2013tya}. 
At this point, recombination of nucleons into alpha particles
provides sufficient energy to unbind ${\sim}10{-}20\%$ of the disc.
The nuclear binding energy liberated in the process is
${\simeq}8.8$~MeV/nucleon. Because the disc material starts to
recombine where the nuclear energy equals the gravitational binding
energy, a characteristic cylindrical radius $\varpi^\ast$ at which the
wind starts is \cite{Fernandez:2013tya}, 
\begin{equation}
  \frac{G M_{\rm disc} m_b}{\varpi^\ast} \simeq 8.8\, {\rm MeV}\ , 
\end{equation}
where $m_b$ is the baryon mass.
These disc ejecta can be massive and are launched around the equatorial
(orbital) plane with characteristic velocities ${\sim}0.1$~c
\cite{Fernandez:2013tya,Fernandez:2015use,Fahlman:2018llv}. 
Magnetohydrodynamics effects can enhance the secular masses and
asymptotic velocities and boost the disc ejection fraction to up to
${\sim}40\%$ \cite{Siegel:2017nub,Fernandez:2018kax}.  
For long-lived remnants, the composition of the secular ejecta depends 
sensitively on the lifetime of the remnant due to neutrino irradiation
\cite{Fernandez:2015use,Radice:2018pdn}.  

From the above discussion it should be clear that several properties of the 
ejecta (and thus of the kilonova) depend sensitively on the remnant,
although these dependencies are not fully quantified yet.
This is further indicated by the fact that some of the broad features 
of synthetic kilonova light curves computed from fiducial NR data show a
correlation with the tidal parameter $\tilde\Lambda$ (and hence the merger outcome) \cite{Radice:2018pdn}.\\

We finally mention the key elements connecting the ejecta and the 
kilonova emission. For a complete discussion see \cite{Metzger:2019zeh,Rosswog:2017sdn}.
The key quantity determining r-process nucleosynthesis in the ejecta
is the electron fraction $Y_e$ \cite{Lippuner:2015gwa,Rosswog:2017sdn}.
If $Y_e \lesssim 0.2$, then the ejecta produce second and third r-process peak
elements with relative abundances close to Solar ones. 
If $Y_e \gtrsim 0.3$, then the material is not sufficiently neutron
rich to produce lanthanides but first r-process peak elements are
produced. A sharp trasition between these two regimes is marked by
$Y_e \simeq 0.25$. 
The $Y_e$ also determines the photon opacity in the
material~\cite{Tanaka:2013ana,Kasen:2013xka}, drastically altering the
timescale and the effective blackbody temperature of the kilonova emission
\cite{Metzger:2019zeh}. High-$Y_e$ ejecta power kilonovae peaking in the
UV/optical bands within a few hours of the merger (blue), 
while low-$Y_e$ ejecta power kilonovae peaking in the infrared over a
timescale of several days (red).

\section{Conclusion}
\label{conc}

It is useful to summarize by focusing on the concrete examples of the two
BNS events observed so far, GW170817
\cite{TheLIGOScientific:2017qsa,Abbott:2018wiz,LIGOScientific:2018mvr,GBM:2017lvd}
and GW190425 \cite{Abbott:2020uma}.

The source of GW170817 has mass $M\simeq 2.73-2.77\Msun$ and mass
ratio up to $q=1.37$ ($1.89$) depending on the low (high) spin prior
utilized in the GW analysis
\cite{TheLIGOScientific:2017qsa,Abbott:2018wiz,LIGOScientific:2018mvr}.
The merger was not observed but the merger frequency can be accurately
predicted from the probability distribution of $\tilde{\Lambda}$ using the NR fits
discussed in Sec.~\ref{sec:mrg}. One finds that the (broad)
distribution of $\tilde\Lambda$ translates into $f_\text{mrg}=1719^{+163}_{-214}$~Hz~\cite{Breschi:2019srl}.
Combining the GW170817 data with the prompt collapse models of
Sec.~\ref{sec:pc}, it is possible to rigorously predict via a
Bayesian analysis that the probability of prompt BH formation is
${\sim}50-70\%$. However, if the constraint
on the maximum mass $M>1.97\Msun$ from pulsar observations is imposed,
the probability significantly decreases below $10\%$. Hence, prompt
collapse in GW170817 is largely disfavoured by the GW 
analysis \cite{Agathos:2019sah}.

A NS remnant would have emitted GWs at the
characteristic frequency $f_2=2932^{+337}_{-409}$~Hz, that can be again estimated from the
$\tilde\Lambda$ posteriors together with the peak GW luminosity
\cite{Zappa:2017xba,Breschi:2019srl}. A sufficiently sensitive network of GW
antennas could have detected the postmerger GW at
$f_2$ with a peak luminosity   
larger than $10^{55}$~erg/s. 
These frequencies and luminosities might be accessible by
improving the design sensitivity of current ground-based GW detectors
by a factor two-to-three or with next-generation detectors~\cite{Clark:2015zxa,Abbott:2017dke,Torres-Rivas:2018svp,Martynov:2019gvu}.

The NR-based GW analysis of the prompt collapse supports 
the mainstream interpretation of the 
electromagnetic counterparts that suggests the formation of a short-lived NS
remnant~\cite{Margalit:2017dij,Bauswein:2017vtn,Shibata:2017xdx,Radice:2017lry,Ruiz:2017due,Rezzolla:2017aly}. 
AT2017gfo, the kilonova counterpart of GW170817,  has both a blue and
a red component, thus suggesting that the ejecta had a broad range of
compositions with at least a fraction being free of lanthanides. 
A fit of AT2017gfo light curves to a semianalytical two-components 
spherical kilonova model indicates the lanthanide poor (rich) blue
(red) component has mass $2.5\times10^{-2}\, \Msun$
($5.0\times10^{-2}\, \Msun$) and velocity
$0.27$c ($0.15$c) \cite{Cowperthwaite:2017dyu,Villar:2017wcc} (see
also Fig.~\ref{fig:dyn_ej_summary}). 
Similar results are obtained using more sophisticated 1D simulations of
radiation transport along spherical shells of mass ejecta
\cite{Tanvir:2017pws,Tanaka:2017qxj}.
The estimated masses are larger than those predicted from NR for the 
dynamical ejecta and the estimated velocities for the blue
component are smaller than
those expected for disc winds \cite{Fahlman:2018llv}.
Note that in Fig.~\ref{fig:dyn_ej_summary} the BNS models fitting the blue component have soft EOS and masses significantly lower than those of GW170817.
Kilonova models with multiple components help in resolving
the tension \cite{Perego:2017wtu} because the faster dynamical ejecta can be 
irradiated by the underlying disc, thus sustaining the
emission \cite{Metzger:2014ila,Lippuner:2017bfm,Kawaguchi:2018ptg}.  
Also, spiral-wave winds 
\cite{Nedora:2019jhl} and/or highly magnetized winds
\cite{Metzger:2018uni,Fernandez:2018kax,Mosta:2020hlh} might contribute in 
filling the gap.

Within this picture, prompt collapse can be tentatively excluded by the
observation of the blue 
kilonova. Under the assumption of an equal-mass merger,
only a small quantity of shock-heated or disk wind ejecta would be present in this case and
it would be inconsistent with the
${\sim}10^{-2}\Msun$ inferred from the data \cite{Margalit:2017dij}.
A long-lived remnant could be excluded based on the
estimated kinetic energy of the observed kilonova and SGRB afterglow,
that are too low for the energy reservoir of a NS remnant at the mass shedding limit.
Note that alternative scenarios based
on the interaction between a relativistic jet and the ejecta exist
\cite{Lazzati:2016yxl,Bromberg:2017crh,Piro:2017ayh}, but they are 
disfavoured due to the insufficient deposition of thermal energy in the
ejecta \cite{Duffell:2018iig}. 

Under the assumption that the merger remnant was a short-lived NS, the NR
models described in Sec.~\ref{sec:pc} and basic arguments led to
estimates of $\Mmax\lesssim2.1-2.3\Msun$
\cite{Margalit:2017dij,Shibata:2017xdx,Rezzolla:2017aly,Ruiz:2017due}. 
Further, using empirical relations between NS radii and the threshold
mass $\Mthr$ for prompt collapse it is possible to tentatively rule
out EOS predicting minimal NS radii ${<}10$~km and radii at $1.6\Msun$
$\lesssim11$~km \cite{Bauswein:2017vtn}. 
Combining the GW data and the phenomenological fit of the disc mass in
Fig.~\ref{fig:disk} also leads to a possible lower bound on the tidal
parameter and thus a stronger constraint on the tidal
parameter $300\lesssim\tilde\Lambda\lesssim800$
\cite{Radice:2017lry,Radice:2018ozg}.

GW190425 is associated to the heaviest BNS source known to date with
$M\simeq 3.2-3.7\Msun$ \cite{Abbott:2020uma}. The mass ratio of
GW190425 can be as high as $q\sim1.25$ ($q\sim2.5$) for low (high)
spin priors. 
Using the NR prompt collapse models presented in Sec.~\ref{sec:pc}, it
is possible to estimated that the probability for the remnant of
GW190425 to have collapsed promptly to a BH is ${\sim}97\%$ \cite{Abbott:2020uma}. 
For an equal mass merger ($q\sim1$), a prompt collapse does not form a
significant disc as discussed in Sec.~\ref{sec:disc}, and thus no bright
electromagnetic counterparts would be expected from this event,
e.g.~\cite{Foley:2020kus}. 
However, the conclusions would be different in the scenario that
GW190425 was produced by an asymmetric binary with $q\gtrsim1.6$.
For large mass ratios, the prompt collapse threshold
significantly decreases and massive neutron-rich discs are
likely~\cite{Bernuzzi:2020txg}. 
One the one hand, the prompt collapse to BH outcome is strenghtened
in the $q\gtrsim1.6$ scenario. On the other hand, a bright and temporally extended
red kilonova, similar to the one expected for BH-NS binaries, would 
have been an expected counterpart
\cite{Radice:2018xqa,Kyutoku:2020xka,Bernuzzi:2020txg}. \\

To conclude, future science with BNS merger observations will
crucially depend on the quantitative characterization of the merger
outcome. 
While numerical-relativity efforts towards physically realistic and
quantitative models for multimessenger analysis are ongoing, the
interplay between theory, simulations and observations appears
necessary to guide these efforts.

\begin{acknowledgements}
  The author thanks M.Breschi, W.Cook, B.Giacomazzo, V.Nedora, A.Perego, D.Radice,
  F.Schianchi, F.Zappa for discussions, comments and help with the
  manuscript, as well as members of  {\tt CoRe} and of the Prometeo
  Virgo group for discussions and inputs.
  The author acknowledges support by the EU H2020 under ERC Starting
  Grant, no.~BinGraSp-714626.  
\end{acknowledgements}

%

\appendix

\section{Numerical-relativity methods}
\label{sec:NRmethods}

Numerical-relativity simulations are based on the 3+1 formalism of
general relativity~\cite{Gourgoulhon:2007ue,Alcubierre:2008,Baumgarte:2010}. 
This appendix schematically summarizes the main physical effects and techniques implemented in
current state-of-art simulations.
\begin{itemize}
\item \emph{Initial data} for circular merger simulation are prepared by solving
the constraint equations of 3+1 general relativity in presence of a
helical Killing vector and under the assumption of a conformally flat metric \cite{Gourgoulhon:2000nn}.
The EOS used for the initial data are polytropes or constructed from the minimum temperature slice of the
EOS table employed for the evolution assuming neutrino-less beta-equilibrium.  
Consistent initial data for circular mergers with NSs with spin are
constructed with an extension of the formalism that is suitable for
a constant rotation velocity of the NS \cite{Tichy:2011gw,Tichy:2012rp}.
\item \emph{The Einstein equations} are then solved with free-evolution
  schemes like BSSNOK \cite{Nakamura:1987zz,Shibata:1995we,Baumgarte:1998te}
  or Z4c \cite{Bona:2003fj,Gundlach:2005eh,Bernuzzi:2009ex,Hilditch:2012fp} based
  on the conformal decomposition of the metric fields. The latter
  scheme (and variation on the original proposal
  \cite{Alic:2013xsa,Dumbser:2017okk,Mewes:2020vic}) incorporates
  improved constraints 
  propagation and damping properties with respect to BSSNOK and is thus
  preferable to BSSNOK in nonvacuum spacetimes.
  Neutron star spacetime evolutions are also performed with the
  generalized harmonic scheme \cite{Duez:2008rb,Lindblom:2005qh}.
\item \emph{Gauge conditions} are chosen as 1+log and Gamma-driver
  shift similarly to binary black hole simulations
  \cite{Brandt:1997tf,Baker:2005vv,Campanelli:2005dd,vanMeter:2006vi,Brugmann:2008zz}. 
  These conditions handle the singularities' formation and 
  movement as moving punctures \cite{Baiotti:2006wm,Thierfelder:2010dv,Dietrich:2014wja}.
\item \emph{General relativistic magnetohydrodynamics} is formulated in conservative
form \cite{Font:2007zz}. Finite volume methods are typically
employed to solve the hydrodynamics. High-order reconstructions or 
shock-capturing finite diffencing schemes proved to be important for
waveform modeling \cite{Bernuzzi:2012ci,Radice:2013hxh,Bernuzzi:2016pie}. 
Magnetohydrodynamics is typically handled using constrained-transport
schemes to control the magnetic-field divergence
\cite{Evans:1988a,Balsara:1999,Balsara:2001,Londrillo:2003qi,DelZanna:2007pk,Giacomazzo:2010bx,Etienne:2011re,Moesta:2013dna,Kiuchi:2014hja}.
The use of the vector potential \cite{Londrillo:2003qi} combined with the Lorentz gauge helps improving numerical stability when using structured meshes \cite{Etienne:2011re}.  
Another method employed to control the magnetic-field divergence in
some simulations is divergence cleaning
\cite{Dedner:2002a,Neilsen:2005rq,Neilsen:2014hha}.
Nonideal (resistive) magnetohydrodynamics schemes have been
formulated in general relativity although their applications to
mergers have been limited to date \cite{Palenzuela:2012my,Palenzuela:2013hu,Dionysopoulou:2015tda}.
\item Approximate \emph{neutrino transport schemes} are based 
  on the moment formalism \cite{Thorne:1981,Shibata:2011kx} and/or
  the leakage scheme
  \cite{vanRiper:1981mko,Ruffert:1995fs,Rosswog:2003rv,Neilsen:2014hha}.
  In the M1 scheme, the moments representation of the Boltzmann 
  equation is truncated at the
  second moment and a closure interpolating between the thin and thick
  regime is imposed to the equations. The compact binaries simulations
  performed so far with this scheme are performed in the gray regime \cite{Foucart:2016rxm,Foucart:2017mbt,Foucart:2018gis}.
  The leakage scheme is an approximation to the transport problem that 
  accounts for changes to the lepton number and for the loss of energy
  due to the emission of neutrinos.
  Momentum transport and diffusion effects are not taken into account
  by the leakage, but free-streaming neutrinos can be additionally
  treated by combining the leakage with the M1 closure scheme 
  \cite{Sekiguchi:2016bjd} or with the M0 scheme \cite{Radice:2018pdn}. 
  The latter is a simplified but computationally efficient scheme free of the 
  radiation shock artifact that plagues the M1 scheme~\cite{Foucart:2018gis}.
\item \emph{Equations of state} models simulated so far
  include Skyrme models with finite-temperature and composition
  dependency, e.g. the LS220 \cite{Lattimer:1991nc} and the 
  SLy4 \cite{Douchin:2001sv,daSilvaSchneider:2017jpg}; 
  relativistic mean field models \cite{Oertel:2016bki} with
  temperature and composition dependencies like the 
  DD2 \cite{Typel:2009sy,Hempel:2009mc} and the SFHo \cite{Steiner:2012rk};  
  and Brueckner-Hartree-Fock extensions to finite temperature like the 
  BLh \cite{Bombaci:2018ksa}.
  Softening effects at extreme densities have been simulated with EOS
  with $\Lambda$-hyperons like the Shen H. et al.~\cite{Shen:1998gq}
  and the BHB$\Lambda\phi$ \cite{Banik:2014qja}, or with
  quark-deconfinement transitions implemented by relativistic
  mean field models, e.g. \cite{Papazoglou:1998vr,Bombaci:2008kb,Chen:2013tfa,Maslov:2018ghi}.
  Large samples of piecewise polytropic EOS
  \cite{Read:2008iy} or cold/beta-equilibrated microphysical with a
  thermal pressure contribution given by a $\Gamma$-law have been
  simulated by various groups. 
  Most of these EOS are compatible with present nuclear constraints
  and the cold, neutrino-less
  $\beta$-equilibrated matter 
  predicts NS maximum masses and radii within the range allowed by
  current astrophysical constraints, including the recent GW
  constraints.
\item \emph{General-relativistic viscous hydrodynamics} schemes have been
  developed recently. One method is the general-relativistic large eddy
  simulations method (GRLES) \cite{Radice:2017zta}. Another method is based on a
  simplified Israel-Stewart formalism of general-relativistic
  shear-viscous hydrodynamics \cite{Shibata:2017xht,Shibata:2017jyf}.
  Both approaches simulate turbulent viscosity by specifying
  an effective shear parameter proportional to the sound speed,  
  $\nu \propto c_s$, that sets the intensity of the turbulence. 
%
\item The \emph{computational domain} is typically covered by a structured
  grid composed of Cartesian overlapping domains (box-in-box) with 2:1
  mesh refinement between parent and child. Evolutions are performed
  with method of lines and the Berger-Oliger algorithm
   with sub-cycling in time and refluxing
   \cite{Berger:1984zza,Berger:1989a}. The computational domain covers
   from the interior  
  of the stars to the radiation zone, with the possibility of moving
  some of the Cartesian boxes to follow the orbital motion. 
  An outer spherical grid composed of multi-patches is sometimes used
  to extend the radiation zone \cite{Ronchi:1996,Pollney:2009yz}.
  Spherical grids are being explored and could help for long-term
  simulations of the postmerger phase \cite{Mewes:2020vic}.
\item \emph{Gravitational waves} are extracted on
  coordinate spheres at large radii 
  using metric (Regge-Wheeler-Zerilli) 
  or curvature (Newmann-Penrose)
  perturbation theory of spherical or axisymmetric spacetimes
  \cite{Moncrief:1974am,Moncrief:1974bis,Teukolsky:1972my}. 
\item \emph{Mass ejecta} are computed as those fluid elements that
  satisfy either the geodesic criterion, $-u_t > 1$ where $u^\mu$ is
  the fluid's 4-velocity, or the Bernoulli
  criterion, $-hu_t > 1$ where $h\geq1$ is the enthalpy, on
  large-radii extraction spheres. 
  Both criteria are approximate and apply to stationary spacetimes. 
  The former criterion assumes
  the ejecta's fluid elements are on ballistic trajectories
  and neglects the fluid’s pressure; the latter is more appropriate
  for steady flow. 
  The geodesic criterion is usually adopted for the fast dynamical 
  ejecta while the Bernoulli one is employed for the winds.
\item \emph{Black-hole horizons} and BH properties are computed using
  apparent horizons \cite{Thornburg:2003sf}.
\end{itemize}

Publicly available NR datasets from merger simulations exist and
will be significantly growing in the next years. Gravitational
waveforms for hundreds of configurations have been released by the 
{\tt CoRe} collaboration\cite{Dietrich:2018phi}, the 
{\tt MPI/Kyoto} group \cite{Kiuchi:2017pte} and the {\tt SXS}
collaboration \cite{Foucart:2018lhe}
on their websites~\footnote{
  {\small \url{http://www.computational-relativity.org/}}\\
  {\small \url{http://www2.yukawa.kyoto-u.ac.jp/~nr_kyoto/SACRA_PUB/catalog.html}}\\
  {\small \url{https://data.black-holes.org/waveforms/index.html}}
} 
Ejecta data from the {\tt CoRe} collaboration \cite{Radice:2018pdn} are available on
\verb#Zenodo# ~\cite{david_radice_2019_3588344}. 
There exists a \verb#Zenodo# community called {\it NRGW open data} 
that hosts a collection of datasets from numerical
relativity and gravitational waves modeling papers:
\begin{center}
\url{https://zenodo.org/communities/nrgw-opendata/}
\end{center}
Data upload and download are open and welcome.

\bibliographystyle{spphys}       

\end{document}